\documentclass[a4paper,twocolumn,11pt,accepted=2023-04-23]{quantumarticle}
\pdfoutput=1

\usepackage[utf8]{inputenc}
\usepackage[english]{babel}
\usepackage[T1]{fontenc}
\usepackage{amsmath}
\usepackage{amssymb}
\usepackage{hyperref}
\usepackage{float}
\usepackage{booktabs}
\usepackage{multirow}
\usepackage{graphicx}
\usepackage{enumitem}
\usepackage{listings}
\usepackage{tikz}
\usepackage{lipsum}

\newcommand{\ket}[1]{\left| #1 \right\rangle}
\newcommand{\state}[3]{\textsuperscript{#1}#2\textsubscript{#3}}

\newcommand{\Yb}{\textsuperscript{171}Yb\textsuperscript{+}}

\newcommand*\sectionname{Section}
\newcommand*\equationname{Equation}

\newcommand{\sectionref}[1]{\sectionname~\ref{#1}}
\newcommand{\appendixref}[1]{\appendixname~\ref{#1}}
\newcommand{\figureref}[1]{\figurename~\ref{#1}}
\newcommand{\tableref}[1]{\tablename~\ref{#1}}
\newcommand{\equationref}[1]{\equationname~\ref{#1}}

\begin{document}

\title{Sample-efficient verification of continuously-parameterized quantum gates for small quantum processors}

\author[1,3]{Ryan Shaffer}
\email{ryan.shaffer@berkeley.edu}
\thanks{Current affiliation: AWS Quantum Technologies, Seattle, WA 98170, USA. Work done prior to joining Amazon.}
\author[1,3]{Hang Ren}
\author[2,3]{Emiliia Dyrenkova}
\author[4]{Christopher G. Yale}
\author[4]{Daniel S. Lobser}
\author[4]{Ashlyn D. Burch}
\author[4,5,6]{Matthew N. H. Chow}
\author[4]{Melissa C. Revelle}
\author[4]{Susan M. Clark}
\author[1,3]{Hartmut H{\"a}ffner}

\affil[1]{Department of Physics, University of California, Berkeley, CA 94720, USA}
\affil[2]{Department of Electrical Engineering and Computer Sciences, University of California, Berkeley, CA 94720, USA}
\affil[3]{Challenge Institute for Quantum Computation, University of California, Berkeley, CA 94720, USA}
\affil[4]{Sandia National Laboratories, Albuquerque, NM 87123, USA}
\affil[5]{Department of Physics and Astronomy, University of New Mexico, Albuquerque, NM 87131, USA}
\affil[6]{Center for Quantum Information and Control, University of New Mexico, Albuquerque, NM 87131, USA}

\begin{abstract}
    Most near-term quantum information processing devices will not be capable of implementing quantum error correction and the associated logical quantum gate set.
    Instead, quantum circuits will be implemented directly using the physical native gate set of the device.
    These native gates often have a parameterization (e.g., rotation angles) which provide the ability to perform a continuous range of operations.
    Verification of the correct operation of these gates across the allowable range of parameters is important for gaining confidence in the reliability of these devices.
    In this work, we demonstrate a procedure for sample-efficient verification of continuously-parameterized quantum gates for small quantum processors of up to approximately 10 qubits.
    This procedure involves generating random sequences of randomly-parameterized layers of gates chosen from the native gate set of the device, and then stochastically compiling an approximate inverse to this sequence such that executing the full sequence on the device should leave the system near its initial state. We show that fidelity estimates made via this technique have a lower variance than fidelity estimates made via cross-entropy benchmarking. This provides an experimentally-relevant advantage in sample efficiency when estimating the fidelity loss to some desired precision. We describe the experimental realization of this technique using continuously-parameterized quantum gate sets on a trapped-ion quantum processor from Sandia QSCOUT and a superconducting quantum processor from IBM~Q,
    and we demonstrate the sample efficiency advantage of this technique both numerically and experimentally.
\end{abstract}

\maketitle

\section{Introduction}\label{sec:introduction}

Verifying the correct operation of quantum computations is an essential step toward building a reliable and scalable quantum information processing device \cite{Gheorghiu2019VerificationApproaches}. Most commonly, quantum computations are broken down into fundamental building blocks known as \textit{quantum gates}, which may include gates such as the well-known Hadamard, Pauli, and CNOT operations. Verifying the behavior of a device's physically-realizable gates, known as a \textit{native gate set}, is a primary area of research in this field. The most complete techniques for gate verification belong to a family of techniques known as \textit{tomography}. Such techniques include quantum state tomography \cite{Walther2005ExperimentalComputing, Cramer2010EfficientTomography}, quantum process tomography \cite{Merkel2013Self-consistentTomography}, and gate set tomography \cite{Blume-Kohout2017DemonstrationTomography, Nielsen2021GateTomography}. Tomographic techniques produce a complete characterization of a quantum operation, which provides a detailed mathematical description of the errors present in the system. However, tomography is extremely resource-intensive, and although techniques exist to improve its scalability somewhat \cite{Shabani2011EfficientSensing, Lanyon2017EfficientSystem}, its cost still typically scales exponentially with qubit count.

In contrast, \textit{benchmarking} techniques for verifying quantum gates are typically resource-efficient and in principle can be scaled to much larger qubit counts than tomographic techniques. These techniques notably include randomized benchmarking (RB) \cite{Emerson2005ScalableOperators, Knill2008RandomizedGates} and more scalable variants such as cycle benchmarking \cite{Erhard2019CharacterizingBenchmarking}, direct RB \cite{Proctor2019DirectDevices}, and mirror RB \cite{Proctor2022ScalableCircuits}, which involve executing randomized circuits which are equivalent to the identity. Additional techniques include as cross-entropy benchmarking (XEB) \cite{Boixo2018CharacterizingDevices} and matchgate benchmarking \cite{Claes2021CharacterBenchmarking, Helsen2022MatchgateGates}, which compare the ideal and experimental output probabilities of random quantum circuits. Benchmarking provides an incomplete characterization of a quantum system, typically producing a small number of values which attempt to characterize the average error rate of particular operations performed by the system. But as the name implies, such techniques are particularly useful when attempting to compare the performance of distinct devices, since they provide metrics which are ostensibly hardware-agnostic. For example, benchmarking techniques may provide an estimate of the average error rate of executing a CNOT gate or of a device's average state preparation and measurement (SPAM) error.

Because many quantum algorithms and especially quantum error correction schemes are expressed in terms of particular fixed sets of gates -- most commonly, the Clifford+T family, which is universal for quantum computation -- much of the benchmarking literature is focused on verifying the operation of these fixed one-qubit and two-qubit gates, or their device-native equivalents. However, near-term quantum devices are unlikely to implement large-scale quantum error correction \cite{Preskill2018QuantumBeyond}, and instead will implement circuits directly using the physical native gate set of the device. These native gates are often not limited to the fixed gate set used by error correction schemes, but rather have a parameterization which provides the ability to perform a continuous range of operations. For example, a single-qubit operation may frequently be parameterized as $R(\theta, \varphi)$, where $\theta$ is the rotation angle and $\varphi$ is the axis of rotation. Multi-qubit operations may also be parameterized. The typical multi-qubit gate for trapped-ion devices, based on the M{\o}lmer-S{\o}rensen interaction \cite{Srensen1999QuantumMotion, Mlmer1999MultiparticleIons}, can be parameterized as $MS(\theta, \varphi)$, where $\theta$ can be interpreted as the effective rotation angle in the multi-qubit space, and $\varphi$ is the effective multi-qubit axis of rotation \cite{Martinez2016CompilingGates}. In many instances, compiling quantum circuits using continuously-parameterized native gate sets can produce more efficient compilations on physical devices than when limited to fixed gate sets \cite{Nebendahl2009OptimalComputing}.

Systematic and efficient verification techniques for continuously-parameterized quantum gates, therefore, are a key ingredient for near-term quantum computers to reliably take advantage of the full scope of physically-realizable operations. Standard RB-based techniques are sample-efficient and often scalable to large system sizes, but these protocols typically have restrictions on the gate sets (e.g., limited to only Clifford gates) and therefore are not generally applicable to this task. A notable exception is a recent proposed variant of mirror RB for universal gate sets \cite{Hines2022DemonstratingSets}, which provides a scalable RB-like protocol for gate sets which may include non-Cliffords (although some restrictions on the gate set remain) and was demonstrated on a 27-qubit device. Previous work has also developed an interleaved RB technique to estimate the fidelity of an arbitrary gate \cite{Chasseur2015CompleteErrors, Chasseur2017HybridGates}, which allows for verification of a particular instance of a parameterized gate, but not across the range of allowed parameters. The XEB protocol, which was used to demonstrate quantum computational supremacy on a 53-qubit device \cite{Arute2019QuantumProcessor}, uses random quantum circuits formed from an arbitrary continuously-parameterized gate set with randomly-chosen parameters. However, because XEB circuits have a broad probability distribution over measurement outcomes, XEB is less sample-efficient than RB, which ideally concentrates all of the probability on a single measurement outcome.

In this work, we discuss the application of the \textit{randomized analog verification} (RAV) technique \cite{Shaffer2021PracticalSimulators} to the task of verifying an arbitrary gate set consisting of continuously-parameterized quantum gates. By concentrating most of the measurement probability on a single outcome (like RB), RAV provides a sample-efficient protocol for verification of gate sets across the range of allowed parameters. As we show in this work, RAV can be practically implemented for quantum processors of up to approximately 10 qubits.
  In \sectionref{sec:methods}, we provide an overview of the RAV technique and compare it to XEB; we describe the stochastic compilation scheme used in constructing the RAV sequences; and we provide details on the experimental setup of the trapped-ion quantum processor, the Quantum Scientific Computing Open User Testbed (QSCOUT) operated by Sandia National Laboratories \cite{Clark2021EngineeringTestbed}, including the technical details of the functionality required to support the continuously-parameterized gate set and large circuit depths used in the RAV sequences.
  In \sectionref{sec:results}, we provide numerical simulations demonstrating the conditions under which we expect RAV to provide a sample efficiency advantage over XEB,
  and we report experimental demonstrations of this efficiency advantage 
    on the QSCOUT trapped-ion device
    and on a superconducting quantum processor from the publicly-available IBM~Q service \cite{IBM2022IBMQuantum}.
  We conclude with additional discussion of these results in \sectionref{sec:discussion}.

\section{Methods}\label{sec:methods}

\subsection{Randomized analog verification for continuously-parameterized quantum gates}\label{sec:verification}

\subsubsection{RAV protocol}

The verification technique introduced in this work is a gate-based adaptation of the \textit{randomized analog verification} (RAV) protocol for analog quantum simulators \cite{Shaffer2021PracticalSimulators}.
When applied to analog quantum simulations, the RAV protocol consists of running randomized analog sequences of subsets of terms of a target Hamiltonian. In particular, a set of unitary operators is chosen consisting of short, discrete time steps of each of the terms of the target Hamiltonian.
A randomly-generated sequence of these operations is then applied, which evolves the system to some arbitrary state.
Next, an approximate inverse of this sequence, generated using the same set of unitary operators by using a stochastic compilation protocol (see \sectionref{sec:stoq}), is applied to the system, which returns it to the initial state with high probability.

Because current gate-based, non-error-corrected quantum computers are realized by carefully tuning the underlying analog interactions to implement quantum gates with the highest fidelity possible, it is natural to adapt the RAV protocol for use in verifying the behavior of gate-based devices with continuously-parameterized native gates. The RAV protocol is described in \figureref{fig:rav-protocol}. The RAV protocol, like XEB, constructs random sequences of \textit{layers}, each of which consists of some fixed number of each of the device's native gates in some randomly-chosen order. Random parameter values are then provided to each of these gates, which allows the protocol to verify the behavior of the device across the range of allowable parameter values for each gate. But unlike XEB, which proceeds by sampling directly from the output of these random sequences, RAV appends a compiled sequence of layers which approximately inverts the initial sequence. Sequences of varying lengths are generated and run on the target device. Finally, an average \textit{error per layer} is extracted from the results.

\begin{figure*}[p]
    \centering
    \includegraphics[width=0.6\textwidth]{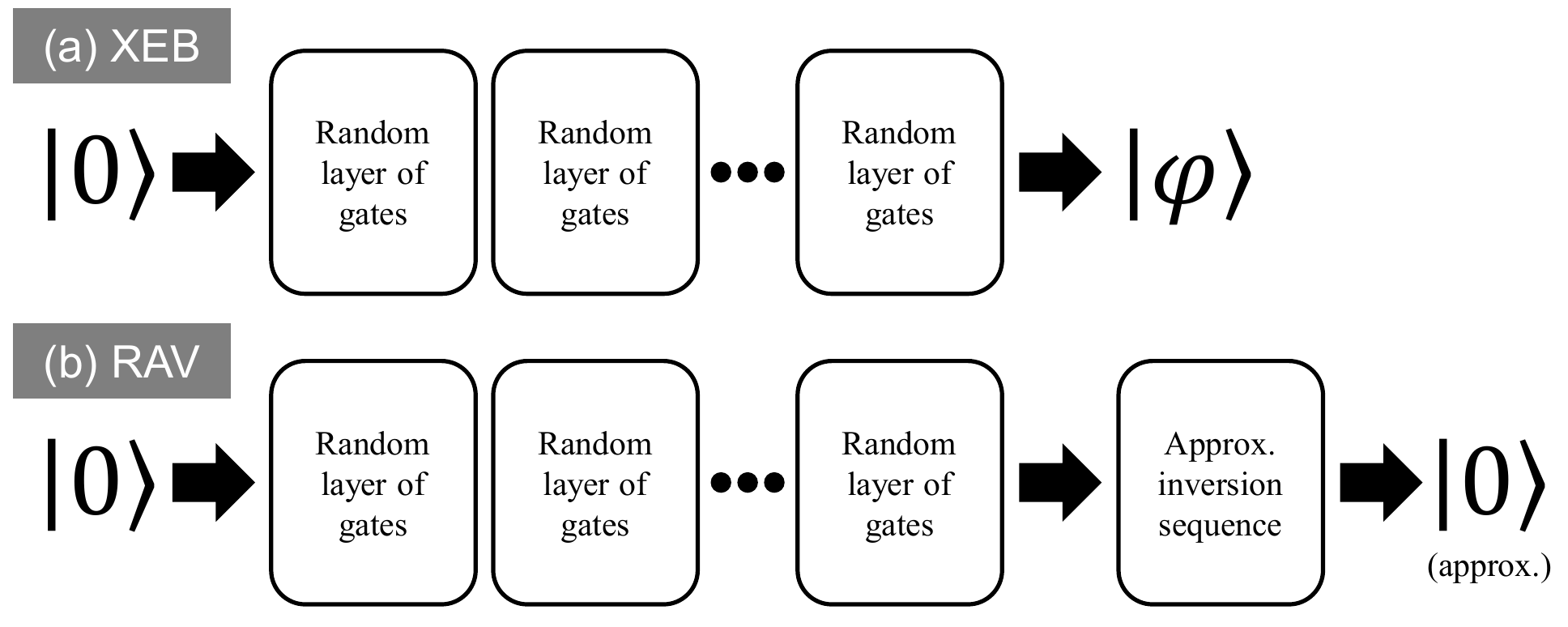}
    \caption{
        Schematics of sequences used in (a) cross-entropy benchmarking (XEB) and (b) randomized analog verification (RAV) protocols for continuously-parameterized quantum gates. The states denoted as $\lvert0\rangle$ can in general be any computational basis state. In XEB, the state denoted as $\lvert\varphi\rangle$ is the ideal result of applying the XEB sequence to the initial state, and is in general far from any computational basis state. Each protocol includes a sequence of randomly-generated layers of native gates with randomly-chosen parameters. In RAV, this is followed by a compiled sequence of layers which is an approximate inverse of the initial sequence and returns the system to the initial state with high probability.
    }
    \label{fig:xeb-rav-schematics}
\end{figure*}

Schematics of the XEB and RAV protocols are displayed in \figureref{fig:xeb-rav-schematics}, which illustrate the fact that the primary difference of RAV from XEB is the inversion sequence which returns the system nearly to the initial state.

\subsubsection{Fidelity estimates using XEB and RAV}

Given a single XEB sequence on an $n$-qubit system, we can approximate the resulting fidelity as
\begin{equation}\label{eq:fxeb}
\hat{F}_{\rm XEB} = \frac{\sum_x P(x) Q(x) - \frac{1}{N}}{\sum_x P(x)^2 - \frac{1}{N}}
\end{equation}
where we have simplified the linear cross-entropy fidelity formula \cite{Boixo2018CharacterizingDevices} for the case of a single circuit.\footnote{The fidelity estimates for XEB and RAV discussed in this section are strictly valid only when averaging over ensembles of circuits; however, we write our expressions in terms of only a single circuit in order to make the analysis more readable.} Here, $P(x)$ represents the classically-computed ideal output probability distribution for the sequence, $Q(x)$ is the observed sample probability of obtaining measurement result $x$, and $N = 2^n$ is the dimension of the system. $\hat{F}_{\rm XEB}$ is constructed such that its observed value for a single circuit might not fall within the range $[0,1]$. But in general, the expected fidelity of the ideal output state (i.e., if $P(x) = Q(x)\ \forall\ x$) is 1, and the expected fidelity of a maximally-mixed output state is 0.

\clearpage
\makeatletter\onecolumngrid@push\makeatother
\begin{figure*}[p]
\centering
\setlength\fboxsep{5mm}
\fbox{%
    \parbox{0.93\textwidth}{%
    \begin{enumerate}[leftmargin=16pt,rightmargin=8pt]
        \item Choose a gate set $G$. Typically this will be the native gate set of the device. Each gate in $G$ is specified as a parameterized unitary (with zero or more parameters), along with the set of allowed target qubits.
        \item Choose a layer design $L_G$ as follows. For each gate in $G$, specify the number of such gates that will appear in each layer, along with allowed ranges for each parameter. A layer is generated from $L_G$ by the following steps:
        \begin{enumerate}[leftmargin=16pt,rightmargin=8pt]
            \item Select each gate the specified number of times.
            \item Choose parameters for each gate uniformly at random from the allowed range.
            \item Choose target qubit(s) for each gate uniformly at random from the allowed set.
            \item Randomly permute the order of the selected gates.
        \end{enumerate}
        \item Generate the RAV sequences as follows. Choose a range of initial layer counts for the RAV sequences $M_0 = (m_{0,\rm min}, \ldots, m_{0,\rm max})$ which can be expected to cover a reasonable range of fidelity loss. For example, $m_{0,\rm min}$ should be small enough to have expected sequence fidelity near 1, and $m_{0,\rm max}$ should be large enough to have expected sequence fidelity near the fully-decayed limit (but not too large). Then, for each $m_0 \in M_0$:
        \begin{enumerate}[leftmargin=16pt,rightmargin=8pt]
            \item Generate a sequence of $m_0$ random layers, each generated independently according to the layer design $L_G$.
            \item Calculate the product of this sequence $U$. Generate an approximate compilation using STOQ (see \sectionref{sec:stoq}), where the target unitary is $U^\dag$ and the instruction set consists of layers generated by $L_G$. The resulting inversion sequence has length $m_{\rm inv}$ and error $\epsilon = 1 - \tfrac{1}{N^2} \left\vert \textrm{Tr} ( V U ) \right\vert^2$, where $V$ is the product of the generated inversion sequence and $N=2^n$ is the dimension of the $n$-qubit system. If necessary, repeat the STOQ compilation until a desired $\epsilon$ is achieved.
            \item Concatenate the initial random sequence and the inversion sequence to produce the final RAV sequence with layer count $m = m_0 + m_{\rm inv}$. This sequence ideally leaves the system in the initial state $x_0$ with probability $P(x_0) = 1-\epsilon$ when averaged over all initial states.
        \end{enumerate}
        \item Run $K$ shots of each RAV sequence, with randomly chosen initial states, and record the probability $Q(x_0)$ of measuring the system to be in the initial state after running the sequence. Calculate $\hat{F}_{\rm RAV}$ for each sequence as specified in \equationref{eq:frav}.
        \item Plot $\hat{F}_{\rm RAV}$ as a function of the total layer count $m$. Assuming negligible state preparation errors, fit the results to an appropriate fidelity loss curve based on the properties of the experimental error, e.g., an exponential decay curve $\hat{F}_{\rm RAV} = \alpha^m$ or a Gaussian curve $\hat{F}_{\rm RAV} = \alpha^{m^2}$ (see \sectionref{sec:rav-fitting-error-rates} for further discussion).
    \end{enumerate}
        }%
}
    \caption{
        Description of the randomized analog verification (RAV) protocol for continuously-parameterized gate sets.
    }
    \label{fig:rav-protocol}
\end{figure*}
\clearpage
\makeatletter\onecolumngrid@pop\makeatother

To derive a formula for the fidelity of a single RAV sequence on an $n$-qubit system, we start with the XEB formula in \equationref{eq:fxeb}. We then note that, by construction of the RAV sequence, $P(x_0) = 1-\epsilon$, where $x_0$ is the initial state (and expected final state) of the RAV sequence and $\epsilon$ is the approximation error of the inversion sequence. After some simplification (see  \appendixref{sec:appendix-rav-fidelity}), we arrive at the following expression for the approximate RAV sequence fidelity:
\begin{equation}\label{eq:frav}
\hat{F}_{\rm RAV}
= \frac{Q(x_0) - \frac{1}{N}}{P(x_0)- \frac{1}{N}}.
\end{equation}

We use the hat on the symbols $\hat{F}_{\rm XEB}$ and $\hat{F}_{\rm RAV}$ to emphasize that they are only estimates of fidelity based on a single circuit instance. To obtain reliable information about the fidelity, these results must be aggregated over many circuit instances. In addition, the use of the linear cross-entropy to estimate $\hat{F}_{\rm XEB}$ as in \equationref{eq:fxeb} is only strictly true in the limit of large circuit depth and when averaged over ensembles of random circuits \cite{Boixo2018CharacterizingDevices}; therefore, the expression for $\hat{F}_{\rm RAV}$ in \equationref{eq:frav} is subject to the same limitation. However, as we demonstrate by numerical simulations (see \figureref{fig:xeb-rav-analysis}), the derived variance of $\hat{F}_{\rm RAV}$ agrees well with observed data even in the regime of small qubit count $2 \le n \le 8$ and moderate circuit depth of $10 \le m \le 30$ layers.

The quantity estimated by $\hat{F}_{\rm XEB}$ and $\hat{F}_{\rm RAV}$ is known as the \textit{depolarization fidelity} \cite{Arute2019QuantumProcessor}. For a circuit whose ideal pure output state is $\lvert \psi \rangle$ and whose execution is subject to purely depolarizing errors, the true mixed output state can be defined in terms of a depolarization parameter $\lambda$ as
\begin{equation}
\label{eq:mixed-state}
\rho_\lambda = (1 - \lambda) \lvert \psi \rangle \langle \psi \rvert + \dfrac{\lambda}{N} I
\end{equation}
where $\lambda \in [0,1]$ is the fraction to which the output state is depolarized. The depolarization fidelity of this state is then defined as $F = 1-\lambda$. This quantity is related to, but distinct from, the typical state fidelity $f = \left(\textrm{Tr}{\sqrt{\sqrt{\rho_\lambda} \lvert \psi \rangle \langle \psi \rvert\sqrt{\rho_\lambda}}}\right)^2$. As derived in \cite{Arute2019QuantumProcessor}, the average depolarization fidelity $\overline{F}$ is related to the average state fidelity $\overline{f}$ as $\overline{f} = \overline{F} + (1 - \overline{F})/N$.
For a fully-depolarized system, the average depolarization fidelity $\overline{F} = 0$, whereas the average state fidelity $\overline{f} = \tfrac{1}{N}$.
We also note that because depolarization fidelity is linearly related to state fidelity, the shape of the loss curve of either quantity will maintain the same character (e.g., exponential or Gaussian).

\subsubsection{Sample efficiency of RAV}\label{sec:rav-sample-efficiency}

Intuitively, we expect that measuring the success of RAV sequences should be more sample-efficient than measuring the success of XEB sequences. This is because RAV requires estimating the output probability $Q(x_0)$ of only the initial state (under the assumptions used to derive \equationref{eq:frav}), whereas XEB requires sampling from the full output probability distribution $Q(x)$. Additionally, in the case where the error is small, we expect the final state of a RAV sequence to be close to a basis state, which minimizes the quantum projection noise associated with the measurement.

More concretely, we can demonstrate this sample efficiency advantage by calculating the statistical variance associated with measurement of $\hat{F}_{\rm RAV}$ and $\hat{F}_{\rm XEB}$ for a single sequence. By making several simplifying assumptions (see \appendixref{sec:appendix-variance} for full details), we can approximate the variance of these quantities as
\begin{align}
    {\rm Var}\big[ \hat{F}_{\rm RAV} \big]
    &\approx 
    \tfrac{1}{K} \left(\tfrac{1}{(1-\epsilon)-\tfrac{1}{N}}\right)^2
    \times \nonumber\\ &\hspace{8mm} \left[ (1-\lambda) (1-\epsilon) + \tfrac{\lambda}{N} \right]
    \times \nonumber\\ &\hspace{8mm} \label{eq:frav-variance}
    \left[1 - (1-\lambda) (1-\epsilon) - \tfrac{\lambda}{N} \right] \\
    {\rm Var}\big[ \hat{F}_{\rm XEB} \big]
    &\approx 
    \tfrac{1}{K} \left( \tfrac{1}{\tfrac{1}{2} - \tfrac{1}{N}} \right)^2 \Big[ \tfrac{1}{2}\left(\tfrac{\lambda}{N}\right)\left(1 - \tfrac{\lambda}{N} \right) +
    \nonumber\\ &\hspace{8mm} \label{eq:fxeb-variance} \tfrac{1}{3}(1-\lambda)\left(1-\tfrac{2\lambda}{N}\right) - \tfrac{1}{4}(1-\lambda)^2 \Big]
\end{align}
where $K$ is the number of independent experimental shots taken for the given sequence
and $\epsilon \ll 1$ is the error in the compiled inversion of the RAV sequence.

\begin{figure*}[p]
    \centering
    \includegraphics[width=0.8\textwidth]{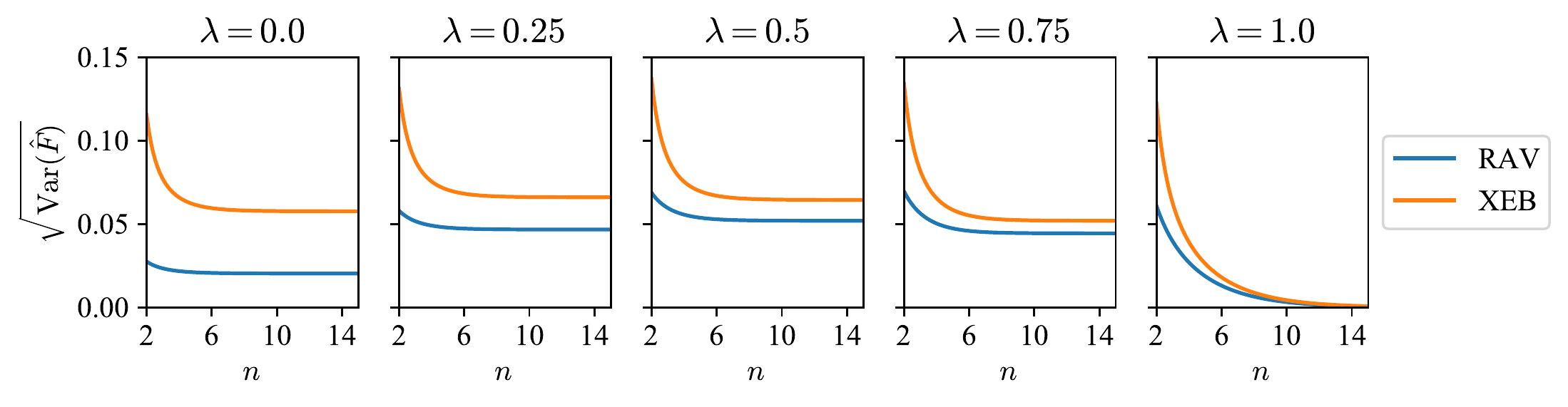}
    \caption{
        Ideal standard deviation of $\hat{F}_{\rm RAV}$ and $\hat{F}_{\rm XEB}$ measurements for $n$-qubit systems as calculated analytically by \equationref{eq:frav-variance} and \equationref{eq:fxeb-variance}, where $\lambda=0$ corresponds to an ideal output state, $\lambda=1$ corresponds to a maximally-mixed output state (see \equationref{eq:mixed-state}), and we assume a RAV inverse approximation error of $\epsilon = 0.04$.
        All plots use $K=100$ independent measurement shots per sequence.
    }
    \label{fig:xeb-rav-ideal-variance}
\end{figure*}

We plot the standard deviation (i.e., $\sqrt{{\rm Var}\big[ \hat{F}_{\rm RAV} \big]}$ and $\sqrt{{\rm Var}\big[ \hat{F}_{\rm XEB} \big]}$) of these fidelity estimates  in \figureref{fig:xeb-rav-ideal-variance} for $2 \le n \le 16$ qubits and $0 \le \lambda \le 1$, assuming $K=100$ and $\epsilon=0.04$. The mean of the RAV and XEB fidelity estimates are in agreement in all cases, but we observe from these plots that RAV fidelity estimates have a lower standard deviation than XEB in all cases, with the advantage tending to shrink as $\lambda$ and $n$ increase.
This implies that fewer RAV repetitions are necessary in order to get an equivalently-precise estimate of the fidelity of a given sequence.

\clearpage
\makeatletter\onecolumngrid@push\makeatother

\begin{figure*}[p]
    \includegraphics[width=1.0\textwidth]{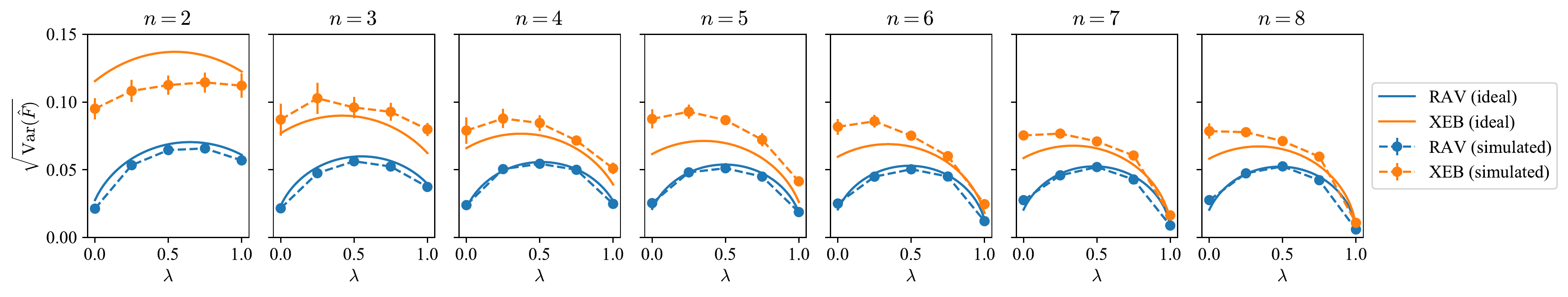}
    \caption{
        Comparison of the precision of fidelity measurements using RAV and XEB at various levels of fidelity loss, where $\lambda=0$ corresponds to an ideal output state and $\lambda=1$ corresponds to a maximally-mixed output state (see \equationref{eq:mixed-state}).
        The ``simulated'' data points, marked by circles, represent the observed standard deviation of 100 simulated measurements of $\hat{F}_{\rm RAV}$ (or $\hat{F}_{\rm XEB}$) for a set of representative RAV (or XEB) sequences on an $n$-qubit system with $10 \le m \le 30$ layers per circuit. Error bars indicate standard error of the mean across the set of sequences; for the RAV data points, the error bars are smaller than the data markers.
        The ``ideal'' solid curves represent the ideal standard deviation of $\hat{F}_{\rm RAV}$ and $\hat{F}_{\rm XEB}$ measurements for the given $n$ and $\lambda$ as calculated analytically by \equationref{eq:frav-variance} and \equationref{eq:fxeb-variance}, assuming a RAV inverse approximation error of $\epsilon = 0.04$.
        All plots use $K=100$ independent measurement shots per sequence.
    }
    \label{fig:xeb-rav-analysis}
\end{figure*}
\clearpage
\makeatletter\onecolumngrid@pop\makeatother

We note that the RAV precision advantage is largest in the regime where $\lambda$ is small, since this is exactly the regime where the final state of a RAV sequence is near a basis state, which minimizes the quantum projection noise associated with the computational basis measurement.

As a further illustration, we perform a simulation of RAV and XEB fidelity measurements across various regimes of fidelity decay assuming a purely depolarizing channel.
We generate representative RAV and XEB sequences for systems with $2 \le n \le 8$ qubits, and we calculate the standard deviation of fidelity measurements on these sequences for $0 \le \lambda \le 1$. These plots, shown in \figureref{fig:xeb-rav-analysis}, show good agreement between the calculated standard deviation for the RAV circuits and the standard deviation extracted from simulations. We observe some quantitative disagreement for the XEB circuits, which is likely because the assumptions behind \equationref{eq:fxeb-variance} are not necessarily valid in the $2 \le n \le 8$ and $10 \le m \le 30$ parameter regime used in these simulations (see \appendixref{sec:appendix-variance}). Nonetheless, we observe qualitative agreement between the calculated and simulated results for the XEB circuits as well as for the RAV circuits, and we observe that the variance of the RAV fidelity estimates is lower than the variance of the XEB fidelity estimates in all cases. This supports the existence of the RAV sample efficiency advantage, even in the absence of the simplifying assumptions used to derive \equationref{eq:frav-variance} and \equationref{eq:fxeb-variance}.

\subsubsection{Fitting RAV fidelity loss curves}\label{sec:rav-fitting-error-rates}

\begin{figure}
    \begin{flushleft}(a)\end{flushleft}
    \vspace{-5mm}
    \includegraphics[width=\columnwidth]{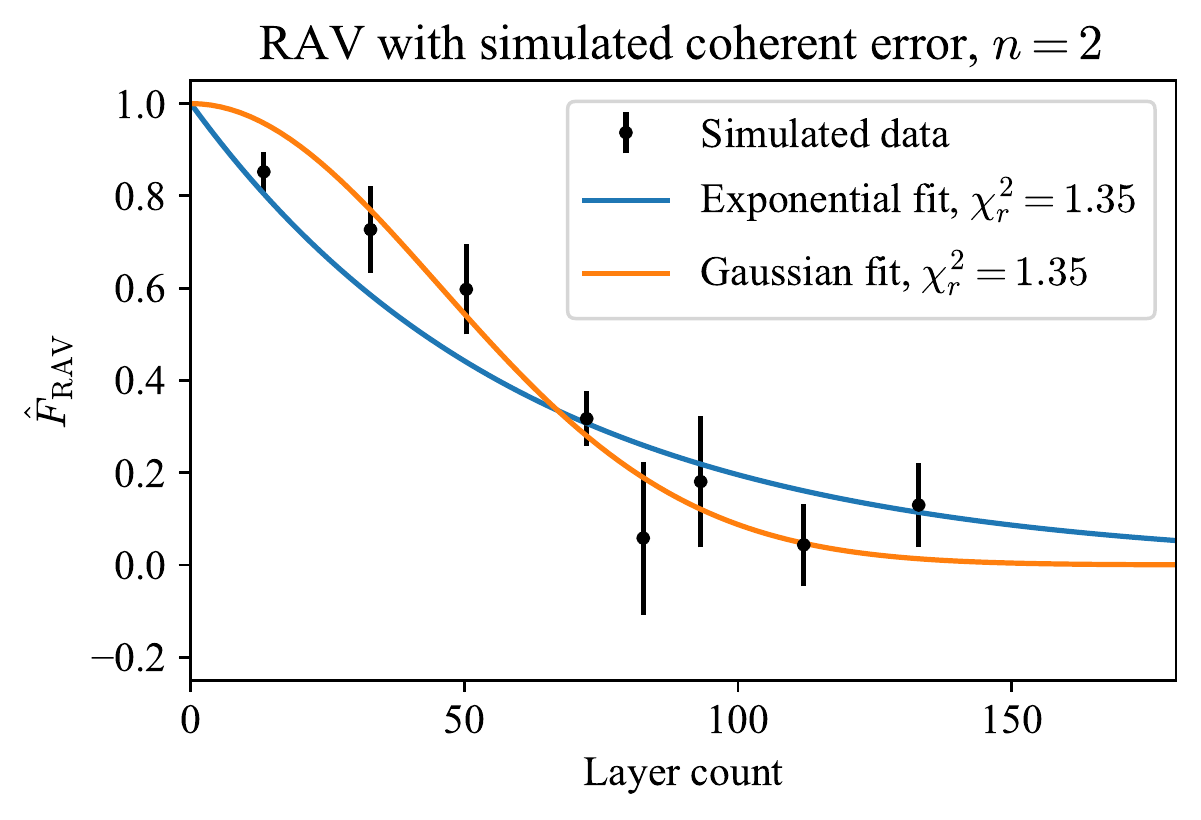}
    \vspace{-5mm}
    \begin{flushleft}(b)\end{flushleft}
    \vspace{-5mm}
    \includegraphics[width=\columnwidth]{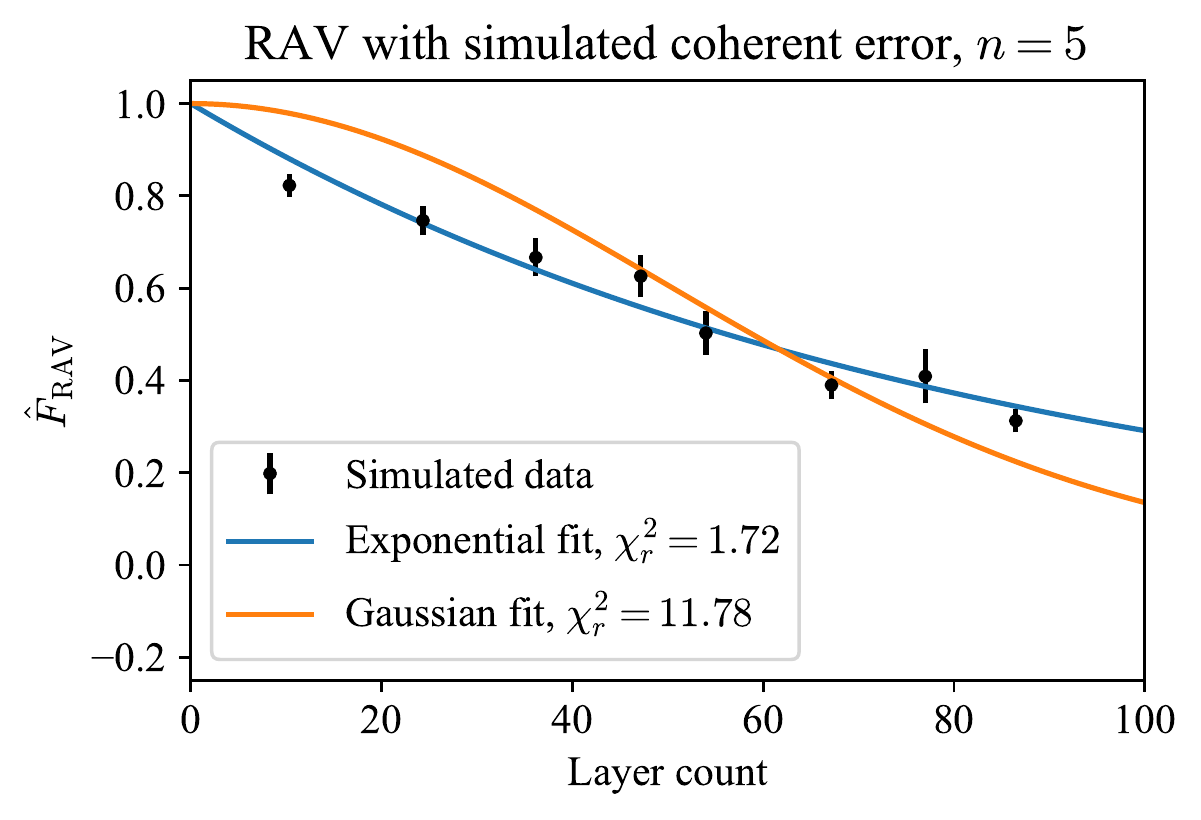}
    \caption{
    Simulated results of RAV sequences under coherent error showing different fidelity loss behavior. Coherent error is implemented in these simulations as a fixed over-rotation of 0.15 radians ($\approx \pi/20$) applied to each physical single-qubit and two-qubit gate. Each data point represents the mean result of six distinct circuits. Error bars represent the standard error of the mean. Fits are a single-parameter exponential curve $\hat{F}_{\rm RAV}=\alpha^{m}$ and a single-parameter Gaussian curve $\hat{F}_{\rm RAV}=\alpha^{m^2}$. Goodness of fit is reported as the reduced chi-squared statistic $\chi^2_r$. (a) Simulated results of 50 RAV sequences on $n=2$ qubits. Sequences are the same as used for experimental demonstration in \sectionref{sec:experimental}.  (b) Simulated results of 50 RAV sequences on $n=5$ qubits. Sequences are the same as used for numerical demonstration in \sectionref{sec:numerical}.
    }
    \label{fig:rav-coherent-error}
\end{figure}

The RAV protocol description (see \figureref{fig:rav-protocol}) does not prescribe a particular fit function, since the shape of the fidelity loss depends on properties of the generated circuits and of the experimental error sources.
Random quantum circuits of large-enough depth (i.e., deep enough that the circuits form approximate 2-designs \cite{Harrow2009Random2-designs}) effectively transform coherent errors into global depolarizing noise \cite{Boixo2018CharacterizingDevices, Dalzell2021RandomNoise}, which would result in a pure single exponential decay. This argument applies to both XEB circuits and RAV circuits, which 
both consist of sequences of random layers.
However, for sequences which are not of sufficient depth, the fidelity loss curve contains additional terms and cannot be accurately modeled as a single exponential decay. Attempting to do so can easily result in an overestimate of per-layer fidelities \cite{Boone2019RandomizedSets,Heinrich2022GeneralCircuits}.
In addition, we note that any real device is subject to a variety of coherent and stochastic error sources which on their own would not act as a depolarizing channel.
So it is not likely that experimental results under various realistic noise sources using relatively small-depth sequences will show good agreement with the idealized results under the assumption of purely depolarizing noise, or that the fidelity estimates extracted under such an assumption will be accurate.

To demonstrate this concretely, we show in \figureref{fig:rav-coherent-error} the results of simulating RAV sequences under a fixed coherent error, which we model as a fixed over-rotation of 0.15 radians ($\approx \pi/20$) applied to each physical single-qubit and two-qubit gate. Sequences are generated using the native gate set described in \sectionref{sec:numerical}. In \figureref{fig:rav-coherent-error}(a), we observe that $\hat{F}_{\rm RAV}$ for these two-qubit RAV sequences appears to follow something in between a Gaussian curve and an exponential curve. In \figureref{fig:rav-coherent-error}(b), for this set of five-qubit RAV sequences, we observe that $\hat{F}_{\rm RAV}$ appears to be fit more closely by an exponential than a Gaussian, but still deviates notably from the ideal exponential decay. These results indicate that the sequences do not satisfy the conditions described above in order to transform the coherent errors into global depolarizing noise. We further note that in an experimental setting, it is likely not feasible to know the shape of the fidelity loss curve in advance, since the properties of the noise are not fully known and may change with time.

For our experimental results in \sectionref{sec:experimental}, we use different fits for the fidelity loss depending on the observed shape of the data. Because this decision is partially dependent on the properties of the particular device, RAV (as the name implies) should be considered a \textit{verification} protocol rather than a \textit{benchmarking} protocol. It provides quantitative metrics about the correctness of a device's operation, but one cannot interpret these metrics in a device-independent way without further assumptions about the types of errors that exist in the device.

\subsection{Stochastic compilation using continuously-parameterized quantum gates}\label{sec:stoq}

We now discuss our technique for compiling the approximate inversion sequence as part of RAV sequence generation. Compiling an inversion sequence for a given random quantum circuit in a non-trivial manner (i.e., other than simply reversing and inverting the original random sequence) into a sequence of continuously-parameterized gates is a difficult problem that in general is infeasible to solve exactly. To produce the inversion portion of the RAV sequences, we introduce a stochastic protocol for approximate quantum unitary compilation, which we abbreviate as STOQ.
We note that this protocol generalizes a similar technique used
   for variational quantum compilation algorithms
      \cite{Khatri2019Quantum-assistedCompiling, Sharma2020NoiseCompiling}.

The process of compilation requires specification of
the unitary operation to be compiled, known as the \textit{target unitary}. This is the $2^n$-dimensional unitary operator $U$ implementing some desired effect on the $n$-qubit system. In the case of RAV, the target unitary is the inverse of the product of the initial randomly-generated sequence of layers.

The set of gates used for the compilation may, in general, be fixed or parameterized.
\textit{Fixed gates}, such as Cliffords, are discrete operations that can be represented as a fixed unitary matrix.
In contrast, \textit{parameterized gates}, such as rotations, are continuous operations that can be represented as a unitary matrix with one or more continuously-variable parameters.
The allowed set of gates for the compilation may then consist of some combination of fixed and parameterized gates. For an $n$-qubit system, the \textit{instruction set} (often called \textit{native gate set}) is a set of fixed gates and/or parameterized gates that represent the fundamental set of operations that can be physically applied to the system.

For a protocol such as RAV, it is desirable that gates occur in the sequence in specific patterns, which we refer to as \textit{layers}. In this case, we can define the instruction set in terms of these layers of fixed and/or parameterized native gates, such that the resulting compilation will be a sequence of layers. Each layer consists of a fixed number of each type of native gate, where each gate is assigned a randomly-chosen parameter value (within some allowed parameter range) and the order of the gates within the layer is also randomly chosen. In the following description of the STOQ algorithm, we refer to the components of an instruction set as \textit{instructions}, where each instruction can be either a single gate or a layer, depending on the definition of the instruction set for the given problem.

Given a target unitary $U$ and an instruction set $G$, then, the goal of an \textit{approximate compilation} protocol is to find a sequence of instructions $\{G_1, \dots, G_M\}$ such that the product $G_M G_{M-1} \cdots G_1$ is as close as possible to $U$ for some reasonable choice of distance metric. Note that this definition does not require any particular closeness of approximation, but it does require that the quality of approximation can be measured. That is, given some appropriate distance metric
  which defines a distance $d$ between
    the sequence product $G_M G_{M-1} \cdots G_1$
    and the target unitary $U$,
an approximate compilation procedure treats $d$ as the value of a cost function to be minimized.

The STOQ protocol for approximate compilation proceeds according to the pseudocode displayed in \figureref{fig:stoq-algorithm}.
Intuitively, the STOQ algorithm can be thought of as
   a randomized exploration of the full set of possible $n$-qubit unitary operators
      (or the subset that can be generated by the instruction set $G$, if $G$ is not universal),
      using a technique known as Markov chain Monte Carlo (MCMC) search \cite{Hastings1970MonteApplications}.
   The algorithm is always initialized with an empty sequence, meaning that it always starts from the identity operator in the search space.
   At each iteration, a random step is proposed, in which an item is either added to or removed from the sequence.
      If this step brings the product of the sequence closer to the target unitary as determined by the cost function, it is accepted;
         otherwise, it is either accepted or rejected with some probability,
         where the probability of accepting such ``bad'' steps decreases with each iteration.
    The algorithm continues until some maximum number of iterations is reached, at which point the final cost can be evaluated and the sequence either kept or discarded, depending on the accuracy requirements of the given problem.

\begin{figure}
\begin{lstlisting}[frame=single]
function StochasticCompilation
(params U, G, num_iterations):
  sequence := []
  beta := 0
  cost := Cost(U, Prod(sequence))
  for i in 1 to num_iterations:
    beta := IncreaseBeta(beta)
    new_sequence := RandomChange(sequence, G)
    new_cost := Cost(U, Prod(new_sequence))
    if Accept(cost, new_cost, beta):
      sequence := new_sequence
      cost := new_cost
  return seq
\end{lstlisting}
    \caption{
        Pseudocode for STOQ stochastic compilation algorithm. 
        The inputs to the algorithm are
            the target unitary \texttt{U},
            the parameterized instruction set \texttt{G},
            and the number of iterations to perform \texttt{num\_iterations}.
        The algorithm is described in \sectionref{sec:stoq}, with additional implementation details provided in \appendixref{sec:appendix-stoq-protocol}.
    }
    \label{fig:stoq-algorithm}
\end{figure}

One critical component of the algorithm is the choice of an appropriate and efficiently-computable cost function. Naturally, the cost function should be a distance measure between the the target unitary $U$ and the unitary $V$ which is the product of the currently-compiled sequence.
   One commonly-used and operationally-relevant choice, used also in variational quantum compilation approaches \cite{Khatri2019Quantum-assistedCompiling, Sharma2020NoiseCompiling},
   is a distance metric defined using the norm of the Hilbert-Schmidt inner product,
      \begin{equation}\label{eq:hilbert-schmidt-distance}
        D_{\textrm{HS}}(U,V)
        = \left\vert \textrm{Tr} ( V^\dag U ) \right\vert
        ,
      \end{equation}
   which is related to the fidelity of a process \cite{Nielsen2002AOperation}.
   We therefore use the cost function
      \begin{equation}\label{eq:cost}
        \texttt{Cost}(U,V)
        = 1 - \frac{1}{2^{n}} D_{\textrm{HS}}(U,V)
        ,
      \end{equation}
   noting that $\texttt{Cost}(U,V)$ ranges from 0 to 1 and vanishes if and only if $U$ and $V$ are equivalent up to a global phase.

Additional technical details and discussion of STOQ are provided in \appendixref{sec:appendix-stoq}, including demonstrations of using STOQ for compilation of sequences to approximately implement time-evolution unitaries, as well as for approximate compilation of randomly-generated unitaries.

\subsection{QSCOUT experimental setup}\label{sec:qscout}

The Quantum Scientific Computing Open User Testbed (QSCOUT) is a quantum processor based on trapped ions housed at Sandia National Laboratories~\cite{Clark2021EngineeringTestbed}.  For the experiments shown here, two \Yb{} ions were used, in which the qubit states are defined by the hyperfine `clock' transition of a \Yb{} ion, \state{2}{S}{1/2}~$|$F=0, $m_F =0\rangle$ ($\ket{0}$) and $|$F=1, $m_F=0\rangle$ ($\ket{1}$).  The ions are controlled via Raman transitions using a pulsed 355\,nm laser in the counter-propagating configuration~\cite{Islam2014BeatProcessing,Hayes2010EntanglementComb}, where one arm is a ``global'' beam that spans all qubits and the other arm consists of up to 32~individual addressing beams, each of which illuminates a single ion~\cite{Debnath2016DemonstrationQubits}. All beams, individual or global, are controlled via the Sandia developed ``Octet'' coherent control hardware with complete two-tone frequency, phase, and amplitude control via rf pulses.  The individual addressing beams are created by a specialized multichannel acousto-optic modulator (AOM) from L3Harris, which divides a single laser beam into 32~beams and propagates each through a separate AOM crystal. 


The single- and two-qubit gates used in the system are all generated via Raman transitions and are parameterized. The single-qubit gates utilize the appropriate individual beam with two tones applied the AOM, generating the necessary transitions in what is known as a co-propagating configuration. Gates about an equatorial axis on the Bloch sphere are physical gates called $R(\theta,\varphi)$, and defined by both $\theta$, which is determined by the duration of the pulse, and $\varphi$, determined by the relative phase of the two tones in the Raman transition. The pulse amplitudes are square-shaped and gapless, meaning there is no ``off'' time between single qubit gates. Fidelities for physical single-qubit gates, $R(\pi/2,0)$ and $R(\pi/2,\pi/2)$ have been estimated to be $99.5\pm0.3\%$ using a variety of techniques including gate set tomography. The gates used for this work do not contain any form of compensation, such as SK1~\cite{Brown2004ArbitrarilySequences}, simplifying the bare gate error analysis and reducing the data acquisition time to limit the effects of drift, especially in the case of significant layer sizes. $R_Z(\theta)$ gates are applied virtually by Octet and seen by the qubits as a cumulative phase shift. 

Two-qubit gates in the system are M\o lmer-S\o rensen (MS) interactions of the form $XX(\theta) = e^{-i\frac{\theta}{2} \sigma_{X} \otimes \sigma_{X}}$ and are also parameterized. They are defined by a desired phase, $\varphi$ and angle, $\theta$.  The MS gate pulses have a Gaussian-shaped amplitude and are composed of one tone on the global beam and two tones each on the individual beams, generating Raman transitions which are detuned symmetrically from a red and blue motional sideband pair.  


Unlike the single-qubit gates, the MS gate has a fixed duration of 200\,$\mu$s, and the rotation angle $\theta$ is determined by the global beam amplitude, accounting for distortions and saturation effects in the global beam amplifier and AOM. In addition, negative rotation angles are generated by changing the relative phase on one of the two ions by $\pi$ radians. For the purposes of this demonstration, all $MS(\theta,\varphi)$ gates had $|\theta| \leq \pi/10$, and for these small values of $\theta$, calibrations suggest deviations in the resultant rotation angle of $\lesssim8\%$. 

Additionally, the MS gates also account for the AC Stark effect through the use of frame rotations, which are virtual Z rotations, applied during the MS gate to cancel phase accumulation from the AC Stark effect. As the global beam and individual beams both contribute to phase shifts caused by the AC Stark shift, the frame rotation also changes depending on the desired $\theta$. These are calibrated to within $\pm 3.5 \times 10^{-3}$ radians, or $\pm0.2$ degrees, for the range of $\theta$ used.

The MS gates are performed in a counter-propagating beam configuration while the single-qubit gates are performed in a co-propagating configuration. Because the relative phase stability between the counter-propagating beams is less stable than that of co-propagating beams, intermixing gates from the two configurations leads to unpredictable phase relationships. To combat this instability, we perform basis transformations on all two-qubit gates. We first surround the MS gate with counter-propagating single-qubit $\pi/2$ gates to transform an $XX$ interaction into a $ZZ$ interaction \cite{Lee2005PhaseGates}. We then further surround those with co-propagating single-qubit gates to bring the interaction back onto an equatorial axis. The desired phase of the MS gate, $\varphi$, is then introduced through the respective phase of these co-propagating single-qubit gates. When including the basis transformation gates, we estimate fidelities for $MS(\pi/2,0)$ to be $97\pm1\%$. 

Due to the nature of the RAV sequences, some extra consideration was needed to deal with non-standard sequences. Low-level pulse data is compressed and stored in a series of lookup tables (LUTs) in Octet's programmable logic for fast readout of data-intensive gate sequences.\footnote{Gate sizes are comparatively larger than other control systems because of the designed flexibility for multi-parameter spline-based modulation \cite{Clark2021EngineeringTestbed}.} The topology of these LUTs and the compression scheme is designed to leverage redundant gate information common to a wide array of quantum circuits. Because of the numerous unique gate calls in RAV sequences, the compression ratio is limited and more LUT storage is required. While storage was increased for certain LUTs using a custom addressing scheme for dense packing of data (not limited to byte-write boundaries), finite memory availability is still a limitation. 

Increasing the LUT storage was supplemented by a compilation technique developed to partially reprogram large segments of the LUTs mid circuit. In this work, the RAV sequences used numerous virtual $R_Z$ gates. Due to their virtual nature, $R_Z$ gates are typically on the order of 10 ns, and continuous streaming of raw data for such short gates exceeds the maximum data throughput supported by the device. The compiler was set up to run recursively, essentially breaking long circuits into smaller pieces based on where the LUT capacity was exceeded. To prevent underflow conditions, partial reprogramming data was placed after gates with long durations by strategic adjustment of the initial boundaries determined by the compiler.

\section{Results}\label{sec:results}

\subsection{Numerical demonstrations}\label{sec:numerical}

\begin{figure}
    \begin{flushleft}(a)\end{flushleft}
    \vspace{-5mm}
    \includegraphics[width=\columnwidth]{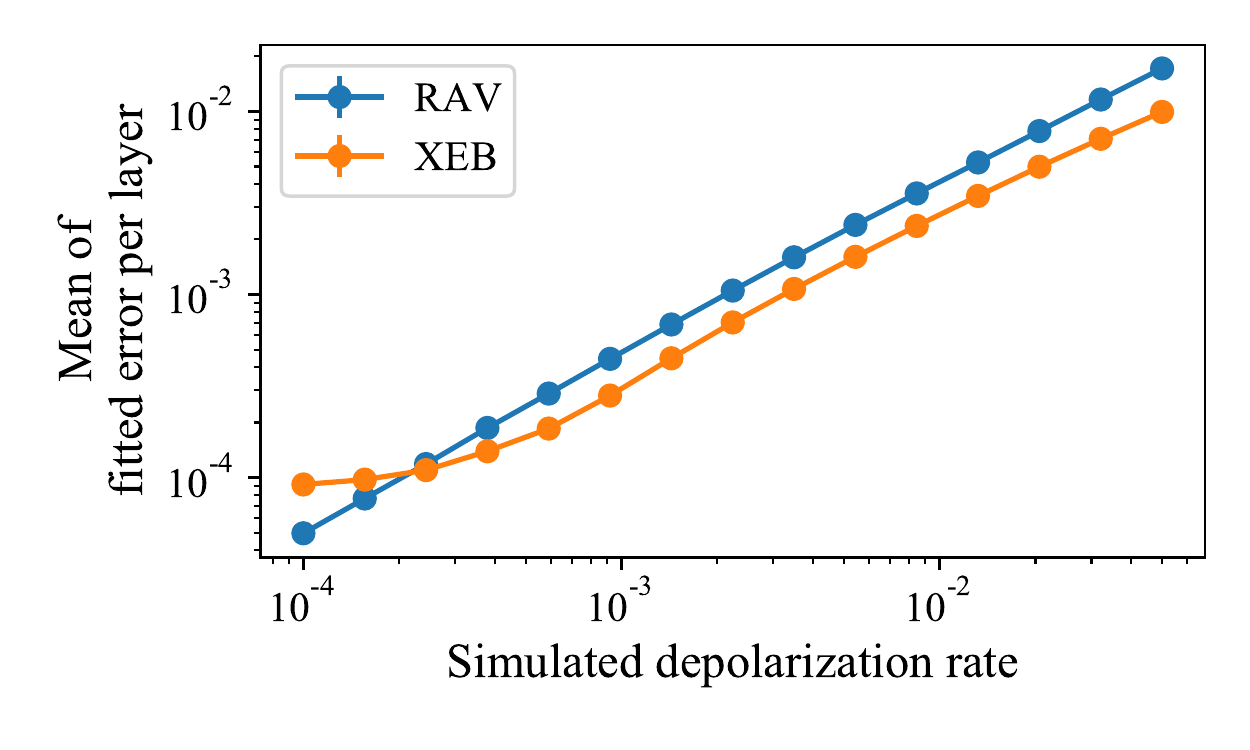}
    \vspace{-5mm}
    \begin{flushleft}(b)\end{flushleft}
    \vspace{-5mm}
    \includegraphics[width=\columnwidth]{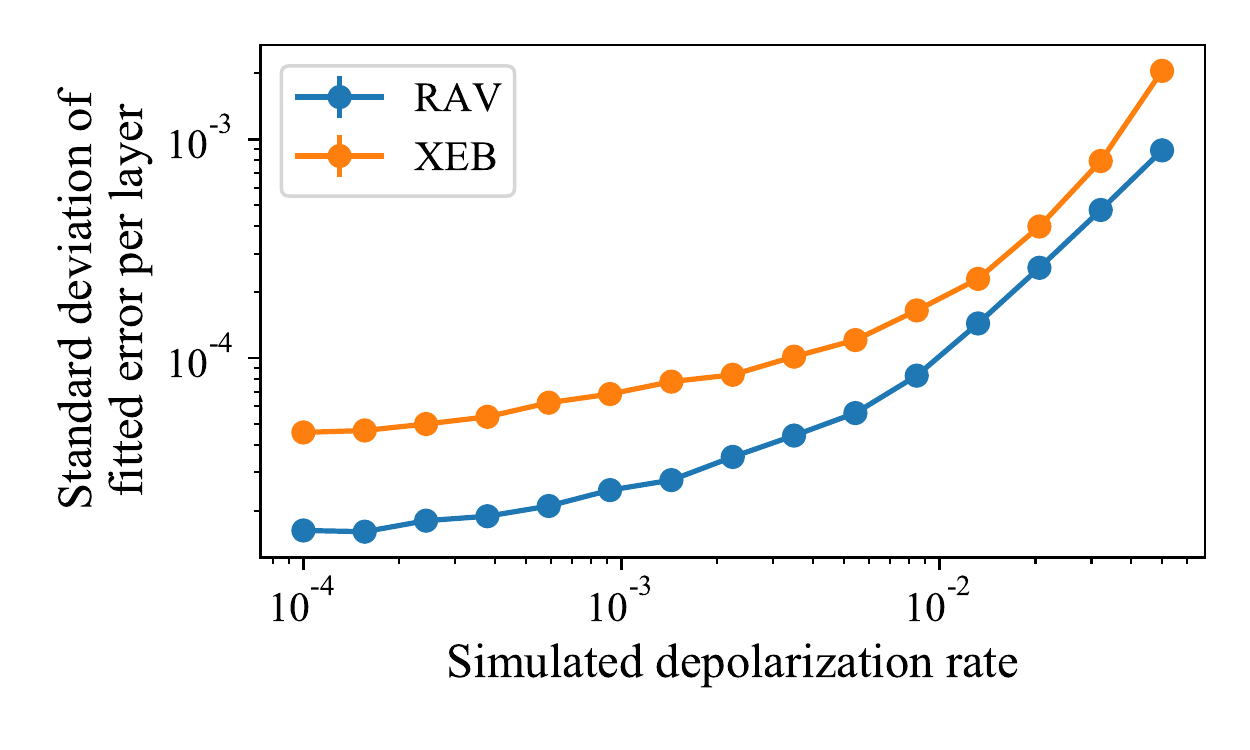}
    \caption{
    Statistics of fitted error rates from simulations of five-qubit RAV and XEB runs, using an exponential fit $\hat{F} = \alpha^m$ to extract the result from each run, and using the quantity $1-\alpha$ as the error rate. Each data point represents the fitted error rate statistics of 100 independent simulations, using 50 RAV or XEB circuits of up to 800 layers with $K=100$ shots per circuit, as described in \sectionref{sec:numerical}. The sets of RAV and XEB sequences are of matching lengths; i.e., for each of the 50 RAV sequences, an XEB sequence was generated with the same number of layers. Error bars, which are smaller than the data markers, represent the standard error of the mean across 10 independent repetitions. Simulated depolarization rate is implemented as the amount of depolarization per $\theta=\pi/2$ rotation via single-qubit $R(\theta, \varphi)$ gate or $\theta=\pi/20$ rotation via two-qubit $MS(\theta, \varphi)$ gate, where the total amount of depolarization is proportional to the rotation angle $\theta$. Simulated $R_Z(\theta)$ gates are unaffected by depolarization rate. (a) Mean fitted error per layer as a function of simulated depolarization rate. (b) Standard deviation of fitted error per layer as a function of simulated depolarization rate.
    }
    \label{fig:numerical}
\end{figure}

To demonstrate the sample efficiency advantage of RAV, we generated 50 RAV and 50 XEB sequences of varying lengths for a five-qubit system. We generated these sequences using a continuously-parameterized native gate set $\{ R(\theta, \varphi), R_Z(\theta), MS(\theta, \varphi) \}$, where these operations are defined in matrix form using the computational basis as follows:
\begin{equation}
    R(\theta, \varphi) 
        = \begin{bmatrix}
           \cos{\frac{\theta}{2}} & -i e^{i\varphi} \sin{\frac{\theta}{2}} \\[2mm]
           i e^{-i\varphi} \sin{\frac{\theta}{2}} & \cos{\frac{\theta}{2}}
        \end{bmatrix}
\end{equation}
\begin{equation}
    R_Z(\theta) 
        = \begin{bmatrix}
           1 & 0 \\[1mm]
           0 & e^{i \theta}
        \end{bmatrix}
\end{equation}
\begin{align}
    MS(\theta, \varphi)
        &= \\ &\hspace{-17mm} \begin{bmatrix}
           \cos{\frac{\theta}{2}} & 0 & 0 & -i e^{-i 2\varphi} \sin{\frac{\theta}{2}} \\[2mm]
           0 & \cos{\frac{\theta}{2}} & -i \sin{\frac{\theta}{2}} & 0 \\[2mm]
           0 & -i \sin{\frac{\theta}{2}} & \cos{\frac{\theta}{2}} & 0 \\[2mm]
           -i e^{i 2\varphi} \sin{\frac{\theta}{2}} & 0 & 0 & \cos{\frac{\theta}{2}}
        \end{bmatrix}\nonumber
\end{align}

We chose this gate set because it aligns most directly with the native gate set of the QSCOUT trapped-ion processor.\footnote{In our implementation of RAV and XEB sequence generation, we used the Python representation of these parameterized operations provided by Sandia at \url{https://gitlab.com/jaqal/qscout-gatemodels/}.} Specifically, $R(\theta, \varphi)$ and $MS(\theta, \varphi)$ are implemented by the QSCOUT device as native one-qubit and two-qubit physical operations, and $R_Z(\theta)$ is implemented by the QSCOUT device as a virtual one-qubit operation which requires no physical interaction with the qubits.

We generated the sets of sequences such that the total layer counts of the RAV and XEB sequences are equivalent. Because RAV sequence lengths are unpredictable due to the stochastic inversion compilation process, we first generated 50 RAV sequences over a range of sequence lengths up to 800 layers. For each RAV sequence, we then generated an XEB sequence of exactly the same length.

Each generated layer consists of three $R^{(i)}(\theta, \varphi)$ gates, three $R_Z^{(i)}(\theta)$ gates, and one $MS^{(i,j)}(\theta, \varphi)$ gate. The target qubit(s) for each gate, represented by the superscript indices, are chosen uniformly at random from the set of all qubits in the system. A layer is constructed by first choosing random parameter values for each of these seven gates. Values for each $\theta$ rotation angle are chosen uniformly at random in the interval $[-\pi/10, \pi/10]$,\footnote{Allowed parameter ranges are chosen with consideration to the performance of the inverse compilation via STOQ. The stochastic search process resembles a stochastic gradient descent, meaning that each generated layer should ideally result in a ``small'' change to the circuit. Empirically, for this gate set, we find that this tends to be satisfied when each rotation angle $\lvert\theta\rvert \ll \pi$. Here we chose the allowed parameter range $\lvert\theta\rvert \le \pi/10$ because this resulted in reasonably good convergence behavior.} and values for each $\varphi$ axis angle are chosen uniformly at random in the interval $[-\pi, \pi]$. After the parameter values are chosen, the order of the gates within the layer is then randomly permuted, which produces the layer that is ultimately used in the sequence.

\figureref{fig:numerical} depicts the mean and standard deviation of error rates per layer obtained via both RAV and XEB for sequences under varying simulated depolarization rates, using an exponential fit $\hat{F} = \alpha^m$ and using the quantity $1-\alpha$ as the error rate. We observe that, over a wide range of depolarization rates, the mean fitted error rates from RAV and XEB are closely aligned, but the error rates estimated by RAV have a significantly smaller standard deviation than those obtained via XEB, indicating that RAV provides more precise information about the overall error rate of these sequences. We also observe that this advantage is more significant in the regime of lower depolarization rates. For a depolarization rate around $10^{-2}$, the RAV error rate estimates are roughly twice as precise as the XEB error rate estimates, whereas around $10^{-4}$, they are roughly three times as precise.

We note that because the RAV sample efficiency advantage comes partially from reduced quantum projection noise due to measurement, we expect that we will find the largest advantage when operating in the early part of the RAV decay curve (which in this simulation is realized by lower depolarization rate). This is the regime where the RAV sequence results remain closest to a basis state, where quantum projection noise is minimized. However, as described previously, RAV continues to have a sample efficiency advantage in the regime of larger error rates because it requires estimating the output probability of only a single state, rather than sampling from the full output probability distribution as required by XEB.

\subsection{Experimental demonstrations}\label{sec:experimental}

To demonstrate the RAV sample efficiency advantage experimentally, we generated RAV and XEB sequences of varying lengths for a 2-qubit system using the same native gate set as in \sectionref{sec:numerical}. In this subsection, we report the results from executing these sequences on quantum processors from QSCOUT and IBM~Q.

\subsubsection{QSCOUT trapped-ion processor}\label{sec:experimental-qscout}

As a first experimental demonstration, we executed 50 RAV and 50 XEB sequences on the two-qubit trapped-ion quantum processor at the Quantum Scientific Computing Open User Testbed (QSCOUT) operated by Sandia National Laboratories \cite{Clark2021EngineeringTestbed}. Details of the experiment are provided in \sectionref{sec:qscout}. We note that this device directly implements the parameterized native gate set $\{ R(\theta, \varphi), R_Z(\theta), MS(\theta, \varphi) \}$ that we used to generate these sequences.

\clearpage
\makeatletter\onecolumngrid@push\makeatother
\begin{figure*}
  \raisebox{-0.95\height}{
    \includegraphics[width=0.29\linewidth]{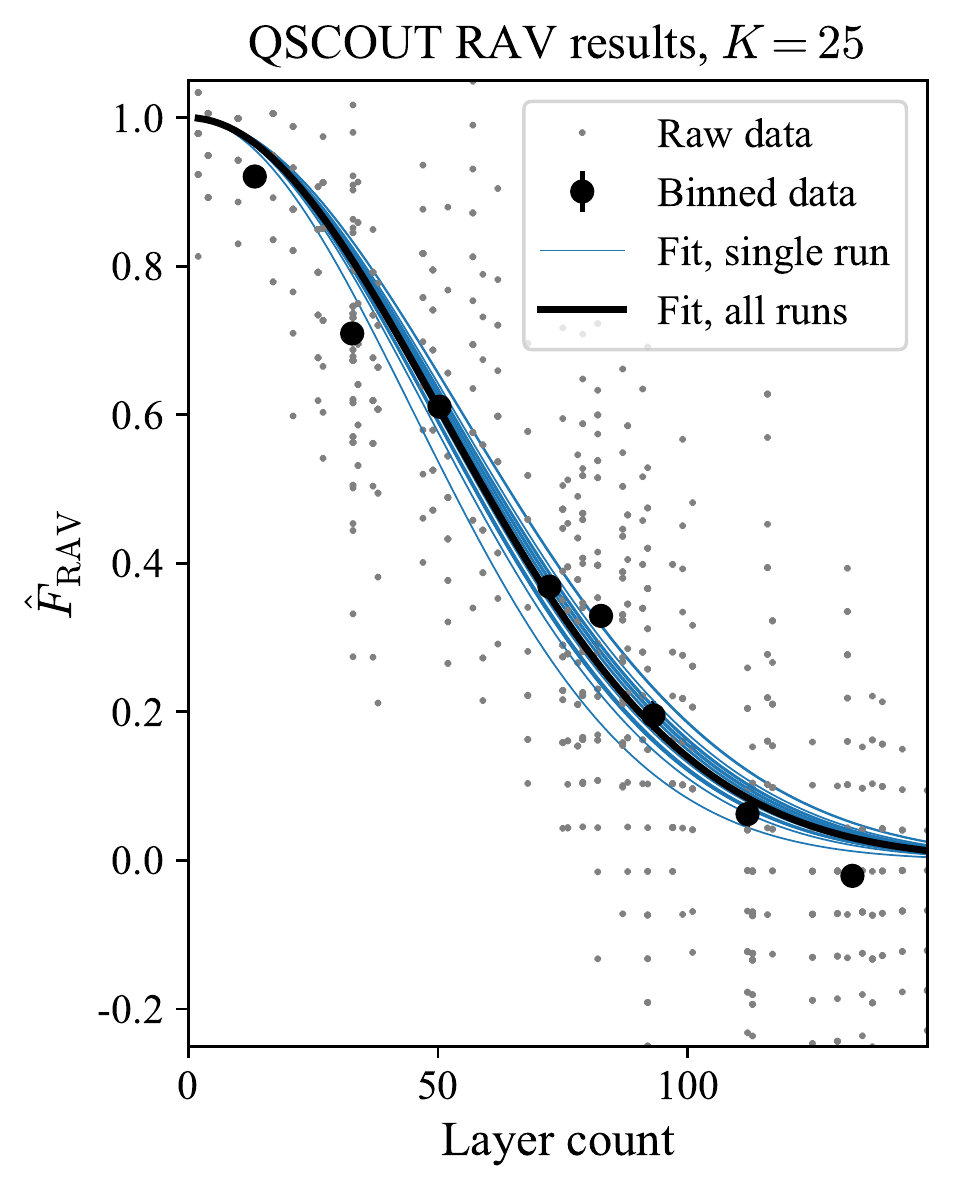}
    }
  \hspace{-.31\linewidth}\mbox{(a)}\hspace{.29\linewidth}
  \raisebox{-0.95\height}{
    \includegraphics[width=0.29\linewidth]{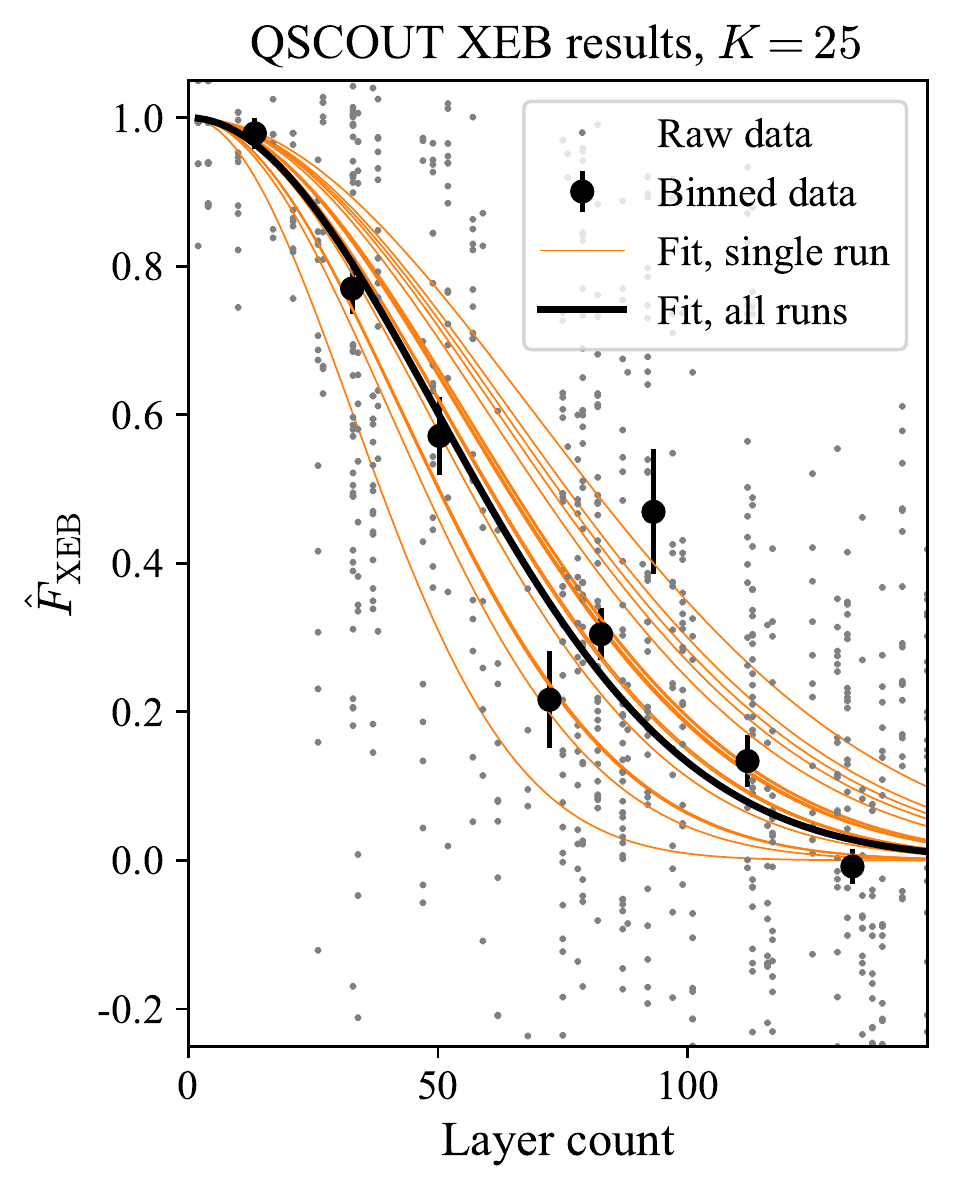}
    }
  \hspace{-.31\linewidth}\mbox{(b)}\hspace{.29\linewidth}
  \raisebox{-0.95\height}{
    \includegraphics[width=0.29\linewidth]{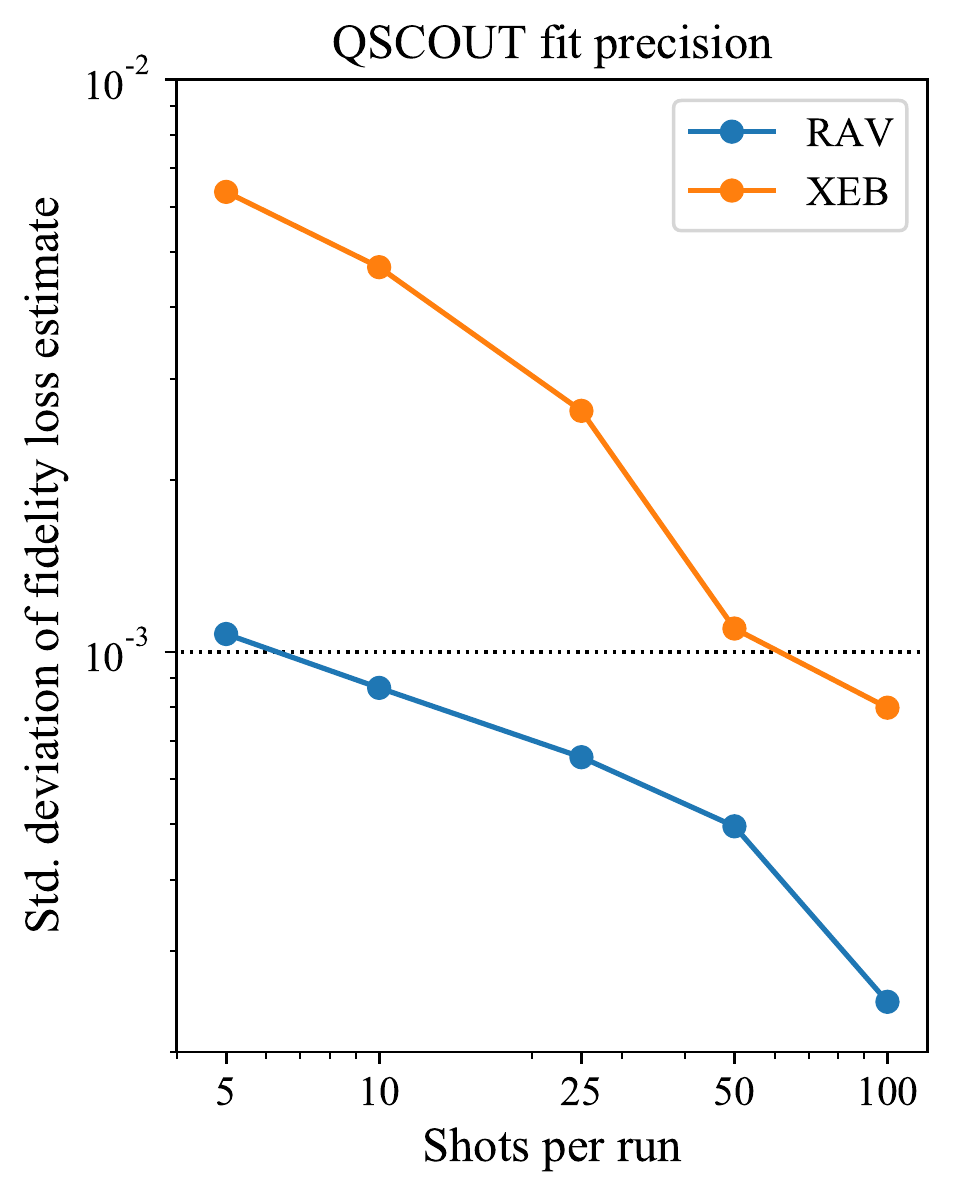}
    }
  \hspace{-.31\linewidth}\mbox{(c)}
  \newline
    \raisebox{1.5cm}{(d)}
  \footnotesize
    \hspace{0.2cm}
\begin{tabular}{@{}c|ccc|ccc|c@{}}
 & \multicolumn{3}{c|}{\textit{RAV: fidelity loss estimate}} & \multicolumn{3}{c|}{\textit{XEB: fidelity loss estimate}} &  \\
 & & & & & & & \\[-1.3em]
     Shots per run
     & Mean $(\times 10^{-2})$ & SD $(\times 10^{-2})$ & $\dfrac{\rm SD}{\rm Mean}$ & Mean $(\times 10^{-2})$ & SD $(\times 10^{-2})$ & $\dfrac{\rm SD}{\rm Mean}$ & 
     $\dfrac{\rm XEB\ \frac{SD}{Mean}}{\rm RAV\ \frac{SD}{Mean}}$ \\[10pt] \hline
$K=5$   & 1.407     & 0.108     & 0.076   & 1.506    & 0.637     & 0.423  & 5.53 \\
$K=10$  & 1.404     & 0.087     & 0.062   & 1.463    & 0.470     & 0.321  & 5.21 \\
$K=25$  & 1.403     & 0.065     & 0.047   & 1.399    & 0.264     & 0.189  & 4.04 \\
$K=50$  & 1.402     & 0.050     & 0.035   & 1.353    & 0.110     & 0.081  & 2.30 \\
$K=100$ & 1.402     & 0.024     & 0.017   & 1.345    & 0.080     & 0.059  & 3.40 \\
\end{tabular}
    \caption{
        Experimental results from QSCOUT for two-qubit RAV and XEB runs. Each run consists of 50 sequences executed $K$ times each.
        (a,b) Results of 20 independent RAV (or XEB) runs using $K=25$ shots per sequence. Raw data points indicate the fidelity estimate $\hat{F}_{\rm RAV}$ (or $\hat{F}_{\rm XEB}$) for $K$ shots of a single sequence, calculated according to \equationref{eq:frav} (or \equationref{eq:fxeb}). Each binned data point (with error bars) represents the mean (and standard error of the mean) of $\hat{F}_{\rm RAV}$ or $\hat{F}_{\rm XEB}$ for a set of six sequences across all 20 runs. Thin curves are single-parameter Gaussian fits $\hat{F}=\alpha^{m^2}$ for the data points from each run, which is chosen based on empirical goodness of fit (see \sectionref{sec:experimental-qscout} for details). The thick curve is the mean of the individual fit curves.
        (c) Standard deviation of the fidelity loss estimate $\sqrt{1-\alpha}$ for RAV and XEB runs for various values of $K$. Smaller standard deviation indicates a more precise estimate of the fidelity loss estimate.
        (d) Fidelity loss estimate statistics for RAV and XEB runs for various values of $K$.
    }
    \label{fig:qscout}
\end{figure*}
\makeatletter\onecolumngrid@pop\makeatother

\figureref{fig:qscout}(a) and \figureref{fig:qscout}(b) show the results of 20 independent runs of the entire set of RAV and XEB sequences, using $K=25$ shots per experiment. The same set of sequences was used for each run, but we performed the 20 independent runs in order to be able to calculate the mean and standard deviation of the estimated fidelity loss obtained from fitting each run to a Gaussian curve $\hat{F} = \alpha^{m^2}$. (For experimental simplicity, 500 executions were performed for each sequence, and we then separated the shot-level results and interpreted them as $500/K$ independent runs of $K$ shots each.)
Based on empirical goodness of fit (see \sectionref{sec:rav-fitting-error-rates}), we choose to fit the data to an Gaussian curve $\hat{F} = \alpha^{m^2}$ (RAV $\chi^2_r = 17.4$, XEB $\chi^2_r = 3.64$) rather than a exponential curve $\hat{F} = \alpha^{m}$ (RAV $\chi^2_r = 64.3$, $\chi^2_r = 19.4$).

It is clear visually in \figureref{fig:qscout}(a) and \figureref{fig:qscout}(b) that the XEB fit curves vary significantly more from the mean than the RAV fit curves.
The standard deviations of the fidelity loss estimates for various values of $K$ are plotted in \figureref{fig:qscout}(c), and the corresponding statistics are tabulated in \figureref{fig:qscout}(d).
As expected, we observe that the the fidelity loss estimates obtained via RAV runs have a significantly smaller relative standard deviation (by a factor of 2.30 to 5.53) than those obtained from XEB runs. Since the standard deviation of the fidelity estimate goes as $1/\sqrt{K}$ (which is supported by the data in \figureref{fig:qscout}(c)), this implies that XEB would require approximately 5 to 30 times as many experimental shots as RAV to produce a fidelity loss estimate for this device with equivalent precision.
For example, \figureref{fig:qscout}(c) illustrates that the RAV $K=5$ runs provide a more precise fidelity loss estimate than the XEB $K=50$ runs.

\clearpage
\makeatletter\onecolumngrid@push\makeatother
\begin{figure*}
  \raisebox{-0.95\height}{
    \includegraphics[width=0.29\linewidth]{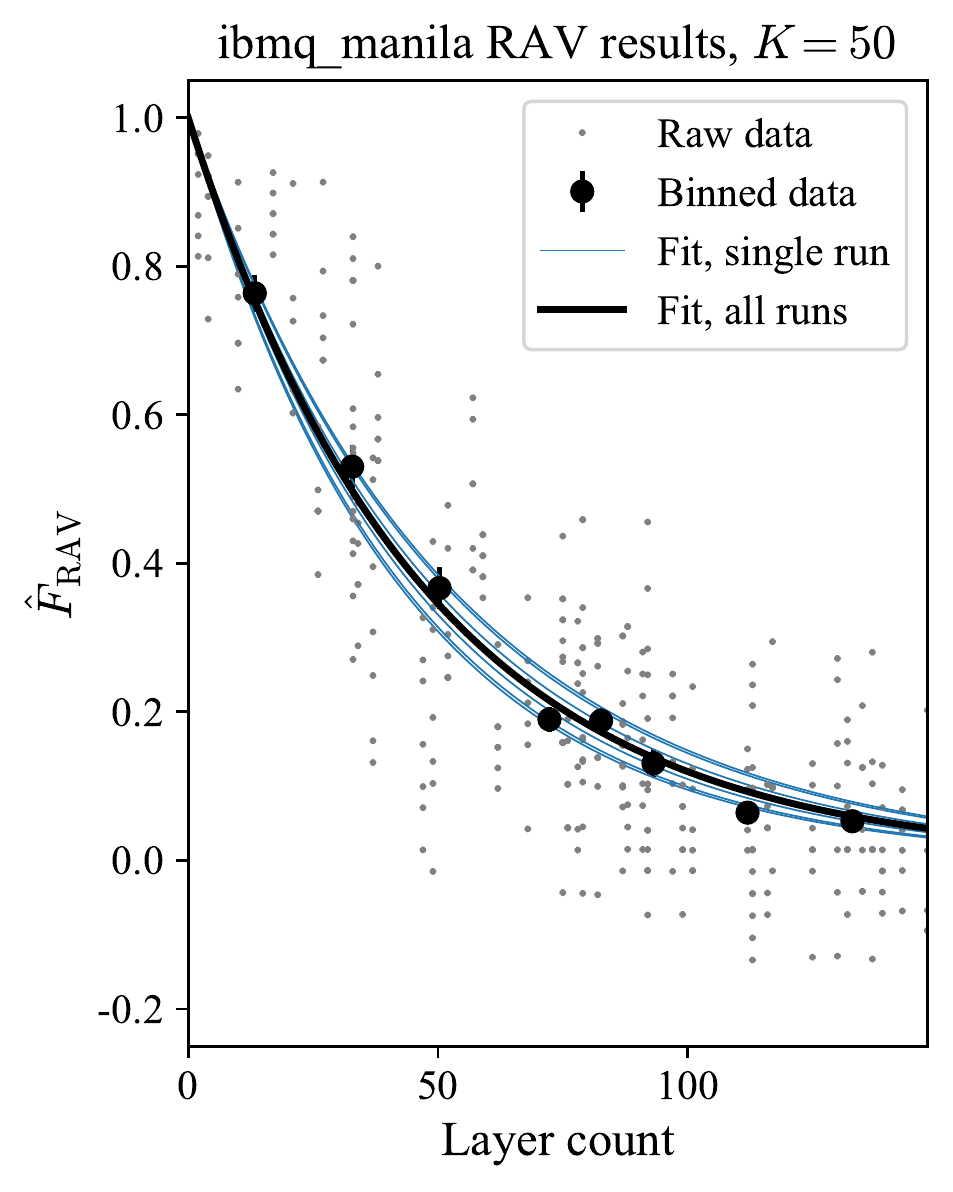}
    }
  \hspace{-.31\linewidth}\mbox{(a)}\hspace{.29\linewidth}
  \raisebox{-0.95\height}{
    \includegraphics[width=0.29\linewidth]{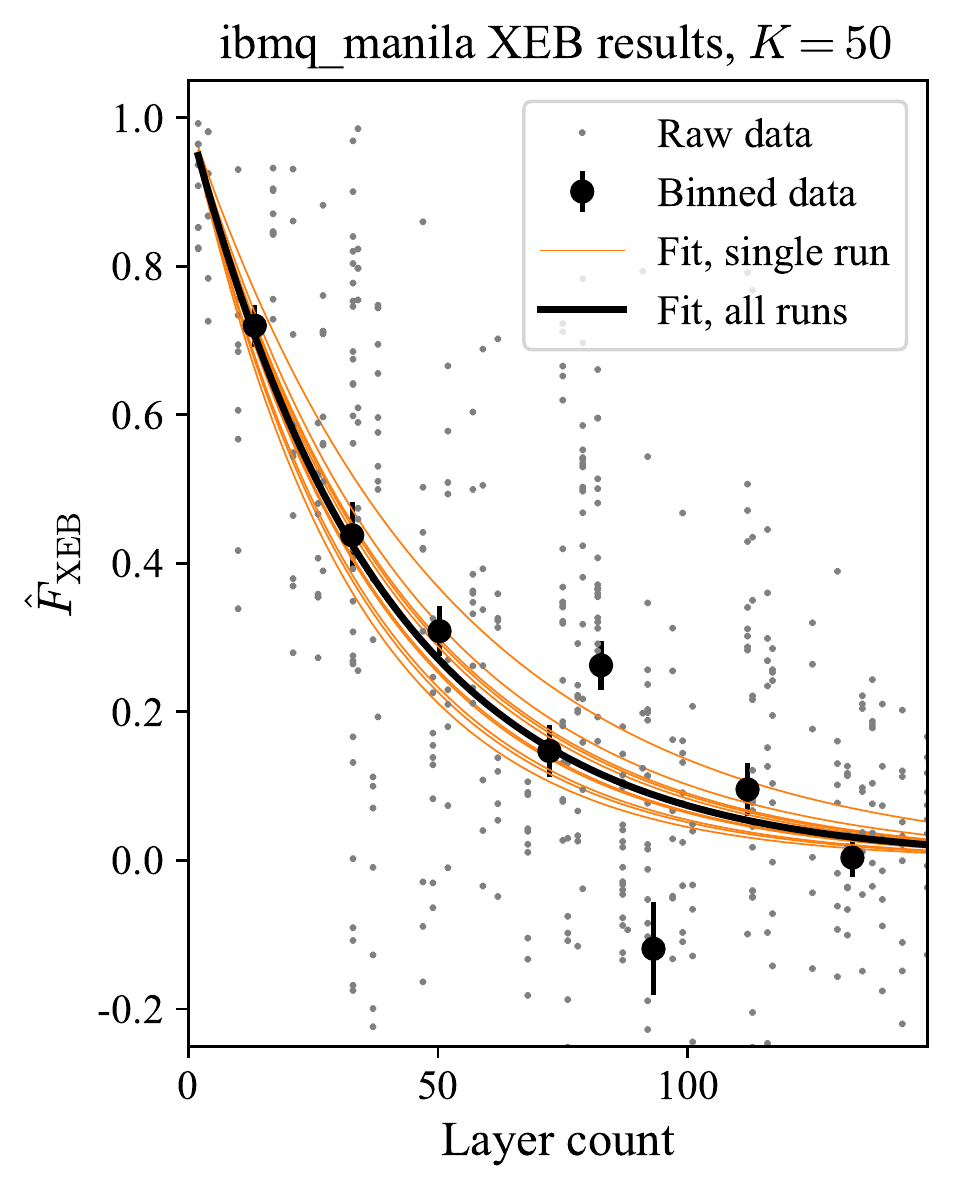}
    }
  \hspace{-.31\linewidth}\mbox{(b)}\hspace{.29\linewidth}
  \raisebox{-0.95\height}{
    \includegraphics[width=0.29\linewidth]{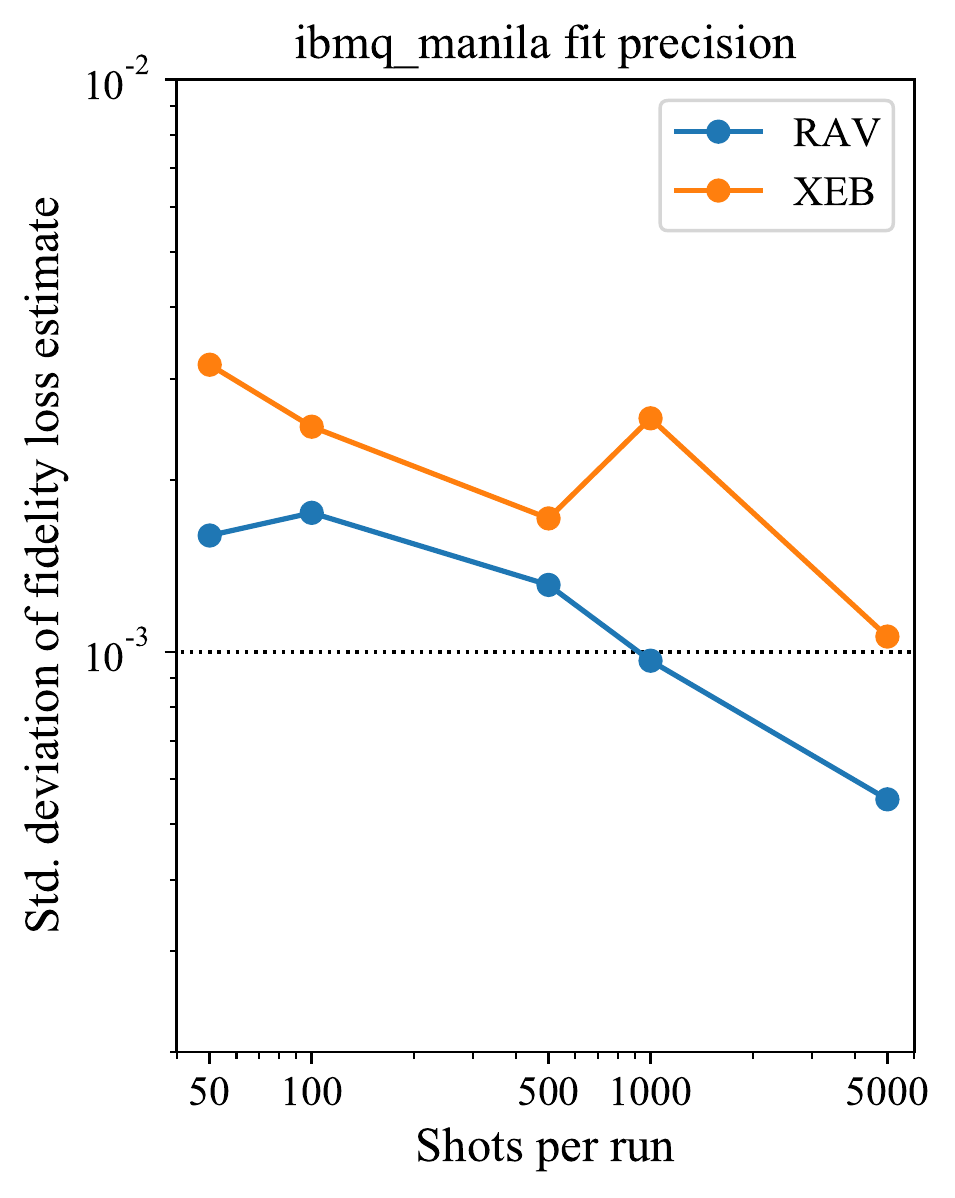}
    }
  \hspace{-.31\linewidth}\mbox{(c)}
  \newline
    \raisebox{1.5cm}{(d)}
    \footnotesize
    \hspace{0.2cm}
\begin{tabular}{@{}c|ccc|ccc|c@{}}
 & \multicolumn{3}{c|}{\textit{RAV: fidelity loss estimate}} & \multicolumn{3}{c|}{\textit{XEB: fidelity loss estimate}} &  \\
 & & & & & & & \\[-1.3em]
     Shots per run
     & Mean $(\times 10^{-2})$ & SD $(\times 10^{-2})$ & $\dfrac{\rm SD}{\rm Mean}$ & Mean $(\times 10^{-2})$ & SD $(\times 10^{-2})$ & $\dfrac{\rm SD}{\rm Mean}$ & 
     $\dfrac{\rm XEB\ \frac{SD}{Mean}}{\rm RAV\ \frac{SD}{Mean}}$ \\[10pt] \hline
$K=50$   & 2.101      & 0.160     & 0.076 & 2.572      & 0.318     & 0.123 & 1.62 \\
$K=100$  & 2.514      & 0.175     & 0.070 & 2.444      & 0.248     & 0.101 & 1.45 \\
$K=500$  & 2.182      & 0.131     & 0.060 & 2.563      & 0.171     & 0.067 & 1.11 \\
$K=1000$ & 2.194      & 0.097     & 0.044 & 2.927      & 0.256     & 0.087 & 1.99 \\
$K=5000$ & 2.047      & 0.055     & 0.027 & 2.434      & 0.106     & 0.044 & 1.62 \\
\end{tabular}
    \caption{
        Experimental results from IBM~Q \texttt{ibmq\_manila} device for two-qubit RAV and XEB runs. Each run consists of 50 sequences executed $K$ times each.
        (a,b) Results of 10 independent RAV (or XEB) runs using $K=50$ shots per sequence. Raw data points indicate the fidelity estimate $\hat{F}_{\rm RAV}$ (or $\hat{F}_{\rm XEB}$) for $K$ shots of a single sequence, calculated according to \equationref{eq:frav} (or \equationref{eq:fxeb}). Each binned data point (with error bars) represents the mean (and standard error of the mean) of $\hat{F}_{\rm RAV}$ or $\hat{F}_{\rm XEB}$ for a set of six sequences across all 10 runs. Thin curves are single-parameter exponential decay fits $\hat{F}=\alpha^m$ for the data points from each individual run, which is chosen based on empirical goodness of fit (see \sectionref{sec:experimental-ibm} for details). The thick curve is the mean of the individual fit curves.
        (c) Standard deviation of the fidelity loss estimate $1-\alpha$ for RAV and XEB runs for various values of $K$. Smaller standard deviation indicates a more precise estimate of the fidelity loss. 
        (d) Fidelity loss estimate statistics for RAV and XEB runs for various values of $K$.
    }
    \label{fig:ibm}
\end{figure*}
\makeatletter\onecolumngrid@pop\makeatother

\subsubsection{IBM~Q superconducting processor}\label{sec:experimental-ibm}

To provide further experimental results, we executed the same 50 RAV and 50 XEB sequences on the publicly-available \texttt{ibmq\_manila} superconducting processor from IBM \cite{IBM2022IBMQuantum}. This device does not directly implement the parameterized native gate set $\{ R(\theta, \varphi), R_Z(\theta), MS(\theta, \varphi) \}$.
To adapt the sequences for this device, we translated the sequences from the original parameterized native gate set into the instruction set that is accepted by the IBM~Q framework. More specifically, $R(\theta,  \varphi)=U(\varphi, \theta-\pi/2, \pi/2-\theta)$ and $MS(\theta, \varphi)=R_Z(-\varphi)^{\otimes2}R_{XX}(\theta)R_Z(\varphi)^{\otimes2}$, where $R_{XX}(\theta)$ is itself a composite gate that can be implemented through multiple native gates. $R_Z(\theta)$ is implementable directly and does not need to be translated.

\figureref{fig:ibm}(a) and \figureref{fig:ibm}(b) show the results of 10 independent runs of the entire set of RAV and XEB sequences, using $K = 50$ shots per experiment. The shapes of the XEB fit curves can be seen to have a larger spread than the RAV fit curves, which again illustrates the smaller uncertainty in estimating fidelity via RAV vs. XEB. 
Based on empirical goodness of fit (see \sectionref{sec:rav-fitting-error-rates}), we choose to fit the data to an exponential curve $\hat{F} = \alpha^{m}$ (RAV $\chi^2_r = 1.46$, XEB $\chi^2_r = 4.76$) rather than a Gaussian curve $\hat{F} = \alpha^{m^2}$ (RAV $\chi^2_r = 24.9$, XEB $\chi^2_r = 17.8$).
The experiment is repeated for varying numbers of shots per sequence, as is shown in  \figureref{fig:ibm}(c). \figureref{fig:ibm}(d) tabulates the specific statistics.\footnote{A total of eight RAV runs and five XEB runs on the IBM~Q device are excluded from the statistics reported in \figureref{fig:ibm} due to abnormally elevated error rates during these particular runs. The qualitative results of the analysis are not affected.}
For some of the experiments, the relative standard deviation of the fidelity loss estimates from the XEB experiments is up to twice that of the RAV experiments, as expected.

Based on the gate error rates published by IBM at the time the experiments were run, we can calculate that the error per layer of our RAV and XEB sequences should be about $1.57 \times 10^{-2}$, which indicates that our experiments overestimated the true error rate. We also note that the IBM~Q results in \figureref{fig:ibm}(c) do not demonstrate the expected scaling of the standard deviation of the fidelity estimate as $1/\sqrt{K}$. We believe that this is because the true error rate of the \texttt{ibmq\_manila} device was fluctuating during the course of the runs, which were spread over the course of several hours. This led directly to increased variance in fitted fidelity loss estimates for some of the experiments; for example, comparing to the QSCOUT results in \figureref{fig:qscout}(c), we clearly see much higher variance in the IBM~Q results despite using the same set of RAV and XEB sequences. Therefore, the variances of our fidelity loss estimates are including the effects of these physical fluctuations in addition to the inherent variance from the RAV and XEB estimates. We expect that runs in which all of the sequences are executed in rapid succession would help to reduce the effect of such fluctuations.

\section{Discussion}\label{sec:discussion}

We note that because of its advantages in sample efficiency, RAV may be particularly useful in the context of frequent calibration runs for devices with continuously-parameterized gates. Minimizing the number of experimental shots per calibration run reduces the downtime of a device due to calibration, and therefore increases its availability for more useful work. Although RAV itself is not a calibration scheme, it can be used to rapidly obtain an estimate of average error over the device's continuously-parameterized gate set (more rapidly than techniques based on XEB, as shown in this work), which could in turn be used as feedback to a calibration technique or as verification of a completed calibration. In this context, the sample efficiency of RAV provides a notable advantage over XEB. We observed from the two-qubit QSCOUT experimental data (see \figureref{fig:qscout}) that achieving a fidelity estimate precision of 0.125\% required only $K=10$ shots of each of the 50 RAV sequences vs. $K=100$ shots of each of the 50 XEB sequences. Each shot took an average of 23~ms experimentally. The corresponding total runtime for a single RAV run is 11.5~s (for 500 total shots), whereas for a single XEB run the total runtime is 115~s (for 5000 total shots). Typical operation of the two-qubit QSCOUT device involves a calibration run approximately every 20 minutes, and so reducing the post-calibration verification step from 115~s to 11.5~s would result in a 10\% increase in the available computational time of the device. We also note that future systems will have larger numbers of qubits and smaller error rates. We expect that both of these aspects will increase the time needed for verification, such that the practical advantage of RAV over XEB in terms of runtime could be substantial.

In this work, we have demonstrated RAV and XEB experimentally on continuously-parameterized gate sets of both a trapped-ion system from Sandia QSCOUT and a superconducting system from IBM~Q. In particular, the gate sets used in this demonstration more closely align with the native physical gates of trapped-ion systems. In QSCOUT, one circuit layer contains a single physical two-qubit gate, where the parameterized $\theta$ is directly associated with physical inputs to that gate; however, on the IBM~Q system, one circuit layer contains two physical two-qubit gates, and the parameterized $\theta$ is instead passed into the surrounding single-qubit gates. As such, the particular RAV construction demonstrated here provides an estimate of the fidelity loss that may provide more direct assessment of physical two-qubit controls for the trapped-ion QSCOUT system, yet still provides an estimate of the fidelity loss for the superconducting IBM~Q system that can be compared to their published error rate. Additionally, RAV could easily be adapted to include gates more closely aligned with the native physical gates of any system, including superconducting systems. 
We also detail the experimental realization of continuously-parameterized two-qubit gates (and relevant calibration bounds to these gates) on the QSCOUT system. We note that additional control-system developments, in the form of extra storage and partial reprogramming, were needed to support the lengthy sequences of unique gate instantiations inherent to verification techniques for continuously-parameterized gate sets. With these capabilities, the RAV and XEB sequences were reasonably practical to experimentally implement on a trapped-ion system such as QSCOUT.

We have demonstrated the generation of RAV sequences on systems of up to $n=8$ qubits. The bottleneck in the RAV sequence generation is the compilation of the approximate inversion sequence via STOQ, which scales poorly with system size (see \appendixref{sec:appendix-stoq-discussion}) and is unlikely to be feasible for $n \gg 10$ qubits. For example, the inversion compilation for each $n=8$ RAV sequence used for \figureref{fig:xeb-rav-analysis} took approximately an hour to generate using a laptop computer. However, this is not an inherent limitation of RAV. If a more efficient technique can be applied to generating the approximate inversion sequence, then RAV sequences could be generated for larger systems. We believe that one promising area of future work would be to adapt the efficient inversion techniques from mirror RB \cite{Proctor2022ScalableCircuits} to the context of RAV and verification of continuously-parameterized gates. For efficient mirroring, this would likely require some restrictions on the construction of layers of the generated RAV circuits, such as requiring that each layer consists only of Clifford gates in order to be efficiently classically simulable.

\section*{Code and Data Availability}

An open-source implementation of RAV sequence generation and the STOQ compilation protocol  \cite{Shaffer2022Rmshaffer/stoq-compiler:V0.2.0}
is freely available at \url{https://github.com/rmshaffer/stoq-compiler}. Data and sequences used to generate the plots in this paper are available upon request from the corresponding author.

\section*{Acknowledgements}

R.S., H.R., E.D., and H.H. acknowledge support from the Challenge Institute for Quantum Computation (CIQC) via the NSF Quantum Leap Challenge Institute (QLCI) program under grant number OMA-2016245,
from the Army Research Office under grant number W911NF-18-1-0170, and from the NSF STAQ project under grant number PHY-1818914. R.S. acknowledges support from the QISE-NET fellowship under NSF award DMR-1747426, as well as from the National Defense Science and Engineering Graduate (NDSEG) fellowship under contract FA9550-11-C-0028
and awarded by the Department of Defense,
Air Force Office of Scientific Research, 32 CFR 168a.

This material was also funded in part by the U.S. Department of Energy, Office of Science, Office of Advanced Scientific Computing Research Quantum Testbed Program. Sandia National Laboratories is a multimission laboratory managed and operated by National Technology \& Engineering Solutions of Sandia, LLC, a wholly owned subsidiary of Honeywell International Inc., for the U.S. Department of Energy's National Nuclear Security Administration under contract DE-NA0003525.  This paper describes objective technical results and analysis. Any subjective views or opinions that might be expressed in the paper do not necessarily represent the views of the U.S. Department of Energy or the United States Government. The United States Government retains and the
publisher, by accepting the article for publication, acknowledges
that the United States Government retains a non-exclusive, paid-up, irrevocable, world-wide license to publish or reproduce the
published form of this manuscript, or allow others to do so, for
United States Government purposes. The Department of Energy will provide public access to these results of federally sponsored research in accordance with the DOE Public Access Plan
(http://energy.gov/downloads/doe-public-access-plan). SAND2023-03170J.

We acknowledge the use of IBM Quantum services for this work. The views expressed are those of the authors, and do not reflect the official policy or position of IBM or the IBM Quantum team.

%
%
\bibliographystyle{quantum}
\bibliography{references}


\clearpage
\widetext
\appendix

\section{Derivation of RAV fidelity estimate}\label{sec:appendix-rav-fidelity}

In this appendix, we derive the formula for the approximate fidelity of a RAV sequence on an $n$-qubit system. We start with the XEB fidelity estimate in \equationref{eq:fxeb}, where $P(x)$ represents the classically-computed ideal output probability distribution for the sequence, $Q(x)$ is the observed sample probability of obtaining measurement result $x$, and $N = 2^n$ is the dimension of the system. The RAV sequence is constructed to return nearly all of the population to the initial state $x_0$. Therefore, we let $P(x_0) = 1-\epsilon$ for some small $\epsilon \ll 1$, which represents the approximation error of the inversion sequence. As a further simplification, we assume that the remaining probability is spread evenly among the remaining states, such that $P(x) = \frac{\epsilon}{N-1}$ for each $x \neq x_0$.

We can then derive the RAV fidelity from the XEB fidelity formula as follows:
\begin{align}
\hat{F}_{\rm RAV}
&= \frac{\sum_x P(x) Q(x) - \frac{1}{N}}{\sum_x P(x)^2 - \frac{1}{N}} \\
&= \frac{P(x_0) Q(x_0) + \sum_{x \neq x_0} P(x) Q(x) - \frac{1}{N}}{P(x_0)^2 + \sum_{x \neq x_0} P(x)^2 - \frac{1}{N}} \\
&= \frac{(1-\epsilon) Q(x_0) + \frac{\epsilon}{N-1} (1-Q(x_0)) - \frac{1}{N}}{(1-\epsilon)^2 + \frac{1}{N-1} \epsilon^2 - \frac{1}{N}} \\
&=
    \frac{N Q(x_0) - 1}{N-1}
  + \epsilon\ \frac{N(N Q(x_0) - 1)}{(N-1)^2}
  + \epsilon^2\ \frac{N^2(N Q(x_0) - 1)}{(N-1)^3}
  + \cdots \\
&=
  \frac{N Q(x_0) - 1}{N-1}
  \left[
    1
    + \frac{N\epsilon}{N-1}
    + \left( \frac{N\epsilon}{N-1} \right)^2
    + \cdots
  \right]
\end{align}
where in the next-to-last step we have performed a Taylor expansion around $\epsilon = 0$.
Then, in the final step we note that the expression inside the square brackets is just a geometric series in $\frac{N\epsilon}{N-1}$, and therefore we have
\begin{align}
    \hat{F}_{\rm RAV} 
    &= \frac{N Q(x_0) - 1}{N-1} \left[ \frac{1}{1-\frac{N\epsilon}{N-1}} \right] \\
    &= \frac{Q(x_0) - \frac{1}{N}}{(1-\epsilon)-\frac{1}{N}}
\end{align}
which, using the fact that $P(x_0) = 1 - \epsilon$, becomes:
\begin{equation}
\hat{F}_{\rm RAV}
= \frac{Q(x_0) - \frac{1}{N}}{P(x_0)- \frac{1}{N}}
\end{equation}

\section{Derivation of variance in RAV and XEB fidelity estimates}\label{sec:appendix-variance}

In this appendix, we derive formulas for the variance of the fidelity estimates resulting from $K$ independent shots of a single RAV or XEB circuit. We start by assuming that the errors in our device are purely depolarizing. Under this assumption, we can represent the output state of any $n$-qubit circuit as a mixture of the circuit's ideal output state
$\lvert \psi \rangle \langle \psi \rvert$ and the maximally-mixed state $\frac{1}{N} I$ (where $N = 2^n$), where $\lambda \in [0,1]$ is the fraction to which the output state is depolarized:
\begin{equation}
\rho_\lambda = (1 - \lambda) \lvert \psi \rangle \langle \psi \rvert + \dfrac{\lambda}{N} I
\end{equation}

We can then define $P(x)$ and $Q_\lambda(x)$ as the probabilities of measuring outcome $x$ when measuring the states $\lvert \psi \rangle \langle \psi \rvert$ and $\rho_\lambda$, respectively, as:
\begin{align}
    P(x) &= \langle x \vert \psi \rangle \langle \psi \vert x \rangle = \big\lvert \langle x \vert \psi \rangle \big\rvert ^2 \\
    Q_{\lambda}(x) &= \langle x \rvert \rho_\lambda \lvert x \rangle = (1-\lambda) P(x) + \frac{\lambda}{N}
\end{align}

We can restate \equationref{eq:fxeb} and \equationref{eq:frav} to define the fidelity estimates $\hat{F}_{\rm RAV}$ and $\hat{F}_{\rm XEB}$ in terms of $P(x)$ and $Q_{\lambda}(x)$ as follows:
\begin{align}
    \label{eq:frav-simple}
    \hat{F}_{\rm RAV}
    &= \frac{Q_\lambda(x_0) - \frac{1}{N}}{P(x_0)- \frac{1}{N}} \\
    \label{eq:fxeb-simple}
    \hat{F}_{\rm XEB}
    &= \frac{\sum_x P(x) Q_\lambda(x) - \frac{1}{N}}{\sum_x P(x)^2 - \frac{1}{N}}
\end{align}

Now, to calculate the expected variance of our these fidelity estimates, we first need to determine their distribution.
To do this, we let $\mathbf{Q}_{\lambda,x} \sim {\rm{Multinomial}}(K, N, Q_\lambda(x))$ be a random variable representing the number of times outcome $x$ is observed when taking $K$ independent shots of a RAV or XEB circuit, where $Q_\lambda(x)$ represents the ``true'' experimental probability distribution of each of $N$ possible outcomes when measuring the state $\rho_\lambda$.

By the properties of the multinomial distribution, then, the variance of $\mathbf{Q}_{\lambda,x}$ is:
\begin{align}
{\rm Var}\big[\mathbf{Q}_{\lambda,x}\big]
&= K Q_\lambda(x) \big[1 - Q_\lambda(x) \big] \\
&= K \left[ (1-\lambda) P(x) + \frac{\lambda}{N} \right] \left[1 - (1-\lambda) P(x) - \frac{\lambda}{N} \right].
\end{align}

We can now construct random variables corresponding to measurements of $\hat{F}_{\rm RAV}$ and $\hat{F}_{\rm XEB}$ by replacing $Q_\lambda(x)$ in \equationref{eq:frav-simple} and \equationref{eq:fxeb-simple} with the scaled random variable $\frac{1}{K} \mathbf{Q}_{\lambda,x}$, which represents the sample probability of observing outcome $x$ when taking $K$ independent shots:
\begin{equation}
\mathbf{\hat{F}_{\rm RAV}} = \frac{\frac{1}{K} \mathbf{Q}_{\lambda,x_0} - \frac{1}{N}}{P(x_0)- \frac{1}{N}}
\qquad
\mathbf{\hat{F}_{\rm XEB}} = \frac{ \sum_x P(x) \frac{1}{K} \mathbf{Q}_{\lambda,x} - \frac{1}{N}}{\sum_x P(x)^2 - \frac{1}{N}}
\end{equation}

Once we have done this, calculating the variance of these random variables is straightforward algebra:
\begin{align}
{\rm Var}\big[ \mathbf{\hat{F}_{\rm RAV}} \big] 
&= \frac{1}{K^2} \left(\frac{1}{P(x_0)-\frac{1}{N}}\right)^2 {\rm Var}\big[\mathbf{Q}_{\lambda,x_0}\big] \\
\label{eq:frav-variance-single-sequence}
&= \frac{1}{K} \left(\frac{1}{P(x_0)-\frac{1}{N}}\right)^2 \left[ (1-\lambda) P(x_0) + \frac{\lambda}{N} \right] \left[1 - (1-\lambda) P(x_0) - \frac{\lambda}{N} \right] \\[5mm]
{\rm Var}\big[ \mathbf{\hat{F}_{\rm XEB}} \big] 
&= {\rm Var}\left[\frac{\frac{1}{K} \sum_x P(x) \mathbf{Q}_{\lambda,x} - \frac{1}{N}}{\sum_x P(x)^2 - \frac{1}{N}}\right] \\
&= \frac{1}{K^2} \frac{\sum_x P(x)^2\ {\rm Var}\left[\mathbf{Q}_{\lambda,x}\right]}{\left(\sum_x P(x)^2 - \frac{1}{N}\right)^2} \\
&= \frac{1}{K^2} \left( \frac{1}{\sum_x P(x)^2 - \frac{1}{N}} \right)^2 \sum_x P(x)^2\ K \left[ (1-\lambda) P(x) + \frac{\lambda}{N} \right] \left[1 - (1-\lambda) P(x) - \frac{\lambda}{N} \right] \\
\label{eq:fxeb-variance-single-sequence}
&= \frac{1}{K} \left( \frac{1}{\sum_x P(x)^2 - \frac{1}{N}} \right)^2
   \Bigg[
     \left(\frac{\lambda}{N}\right)\left(1 - \frac{\lambda}{N} \right) \sum_x P(x)^2
     + (1-\lambda)\left(1-\frac{2\lambda}{N}\right) \sum_x P(x)^3    \nonumber\\&\hspace{50mm} 
     - (1-\lambda)^2 \sum_x P(x)^4
   \Bigg]
\end{align}

The above formulas are sufficient to predict the variance in fidelity measurements for a particular circuit, assuming we know the ideal probabilities $P(x)$ of measuring the circuit output. But of course, these probabilities will be different for every circuit. To make more general predictions, we need to make additional assumptions.

For RAV, the sequences have been explicitly constructed such that most of the population is returned to the initial state $x_0$. Therefore, we assume that $P(x_0) \approx 1-\epsilon$ for some small $\epsilon \ll 1$, which represents the approximation error of the inversion sequence. Substituting this into \equationref{eq:frav-variance-single-sequence} gives
\begin{equation}
    {\rm Var}\big[ \mathbf{\hat{F}_{\rm RAV}} \big] \approx 
    \frac{1}{K} \left(\frac{1}{(1-\epsilon)-\frac{1}{N}}\right)^2 \left[ (1-\lambda) (1-\epsilon) + \frac{\lambda}{N} \right] \left[1 - (1-\lambda) (1-\epsilon) - \frac{\lambda}{N} \right].
\end{equation}

For XEB, it is known that for ensembles of random circuits of large-enough depth on systems with tens of qubits, the distribution of ideal output state probabilities can be well-approximated by a Porter-Thomas distribution \cite{Boixo2018CharacterizingDevices}, in which the probabilities follow an exponential distribution ${\rm Pr}(Np) = N e^{-Np}$.
By the properties of the exponential distribution, then, we have $\sum_x P(x)^k \approx \frac{1}{k}$, and substituting this into \equationref{eq:fxeb-variance-single-sequence} gives
\begin{equation}
\label{eq:fxeb-variance-single-sequence-appendix}
    {\rm Var}\big[ \mathbf{\hat{F}_{\rm XEB}} \big] \approx 
    \frac{1}{K} \left( \frac{1}{\frac{1}{2} - \frac{1}{N}} \right)^2 \left[ \frac{1}{2}\left(\frac{\lambda}{N}\right)\left(1 - \frac{\lambda}{N} \right) + \frac{1}{3}(1-\lambda)\left(1-\frac{2\lambda}{N}\right) - \frac{1}{4}(1-\lambda)^2 \right].
\end{equation}
It is important to note that in \figureref{fig:xeb-rav-analysis} we are working with simulations of sequences with $n \le 8$ qubits and $10 \le m \le 30$ layers, and so the Porter-Thomas assumption is not necessarily valid in this regime. This is the most likely explanation for the systematic discrepancies between this estimate of the XEB variance and the observed XEB variance in our simulations.
We also note that the convergence to the Porter-Thomas distribution applies only to ensembles of circuits, since a fixed circuit will have a particular distribution that does not ``converge'' toward anything. The expression in \equationref{eq:fxeb-variance-single-sequence-appendix} should therefore be thought of as an ``expected'' variance over all XEB circuits, rather than a variance for a particular fixed circuit.

\section{Approximate unitary compilation via stochastic search (STOQ)}\label{sec:appendix-stoq}

In this appendix, we supplement the main text with additional implementation details of STOQ and examples of its use.
    We report results of applying STOQ to Hamiltonian time-evolution unitaries, where we compare its performance to existing methods on various metrics.
    We also demonstrate the use of STOQ to approximately compile gate sequences for randomly-generated unitaries, although as one would expect, its performance scales poorly with the required circuit depth.
    We conclude with additional discussion of STOQ, including its features and limitations.

\subsection{Background}\label{sec:appendix-stoq-background}

A critical prerequisite to executing any algorithm on a physical quantum computer is the process commonly known as quantum compilation.
   One of the primary tasks of quantum compilation is the conversion of a target unitary operation into a sequence of quantum gates that are native to the physical device being used \cite{Barenco1995ElementaryComputation, Cybenko2001ReducingOperations, Harrow2002EfficientGates, Javadiabhari2015ScaffCC:Programs}.
   Because unitary operators belong to a continuous space, such compilation in general results in gate sequences which are only approximately equivalent to the target unitary.
   For example, one of the earliest quantum compilation techniques, the Solovay-Kitaev method \cite{Kitaev1997QuantumCorrection}, compiles gate sequences that differ from the target unitary by an amount that can be made as small as desired.

Traditional compilation, both in the classical and quantum realms, is most often a deterministic process, using rules and heuristics to efficiently synthesize a desired program from
   the native assembly instructions (in classical compilation)
   or native physical gates (in quantum compilation).
But in some cases, adding stochasticity to the compilation process has been shown to produce advantages in the resulting program.
   In classical compilation, a technique known as stochastic superoptimization \cite{Schkufza2013StochasticSuperoptimization} has been shown in certain cases to produce significantly shorter programs than the best-in-class compilers and optimizers.
   In quantum compilation, techniques such as randomized compiling \cite{Wallman2016NoiseCompiling} have been demonstrated to improve noise resilience by randomizing errors that occur during program execution.

In the field of quantum compilation, special attention has been paid to compilation of unitaries which result from the time-evolution of physically-realizable Hamiltonians.
The compiled sequences in these cases can be executed to perform what is known as ``Hamiltonian simulation'', or more broadly, ``quantum simulation''.
Such approaches are of special interest in fields such as quantum chemistry, where it is desirable to use a quantum computer to simulate the dynamics of physical systems.
Common approaches to this problem include
   product formula techniques such as the Suzuki-Trotter decomposition \cite{Hatano2005FindingOrders}
   and qubitization \cite{Low2019HamiltonianQubitization},
which deterministically compile the time-evolution unitary for a given Hamiltonian into a sequence of quantum gates.

Approaches involving stochasticity have recently been shown to be advantageous in some cases.
   Adding randomization to the
    Suzuki-Trotter decomposition \cite{Childs2019FasterRandomization}
    creates approximate compilations that are better both theoretically and empirically.
   A stochastic compilation protocol known as
    QDRIFT \cite{Campbell2019RandomSimulation},
    where gate probabilities are weighted according to the strength of each term in the Hamiltonian rather than using a product formula directly,
    has been shown to produce much more efficient compilations in many cases.
   An interpolation of these two methods \cite{Ouyang2020CompilationSparsification}
    has also been proposed, which takes some of the advantages of each method.
The efficiency of these compilation methods is generally independent of system size when applied to problems involving sparse Hamiltonians.

However, these specialized methods cannot be applied to general-purpose compilation tasks, which is where we focus specifically here.
   In \sectionref{sec:stoq} of the main text, a stochastic approximate quantum unitary compilation procedure, abbreviated as STOQ, is described in detail.
   The STOQ protocol was originally developed as part of a verification scheme for analog quantum simulators called \textit{randomized analog verification} (RAV) \cite{Shaffer2021PracticalSimulators}.
   And in this work, an adaptation of RAV for gate-based quantum devices is summarized in \sectionref{sec:verification}, demonstrated numerically in \sectionref{sec:numerical}, and demonstrated experimentally in \sectionref{sec:experimental}. This gate-based version of RAV also uses STOQ to compile the approximate inversion portion of each sequence.

\subsection{Additional implementation details}\label{sec:appendix-stoq-protocol}

This section fills in a few important details of the STOQ protocol implementation outlined in \sectionref{sec:stoq},
    specifically referring to the pseudocode representation in 
    \figureref{fig:stoq-algorithm}.
    
The compiled sequence is stored in the \texttt{sequence} variable,
    which is initially empty.
    The \texttt{RandomChange} function returns a modified sequence on each iteration, either
        by adding a randomly-drawn instruction to the sequence
            from the parameterized instruction set \texttt{G}
            with randomly-generated parameter values,
        or by removing an instruction from the sequence.
    The \texttt{Prod} function calculates the unitary that represents the product of all of the operations in the sequence,
    and the \texttt{Cost} function is implemented as described in \equationref{eq:cost}.

The variable \texttt{beta} is used as an annealing parameter for the compilation process.
    The function \texttt{IncreaseBeta} returns a slightly increased value of \texttt{beta} on each iteration.
    Defining the annealing parameter as
            $\beta = \texttt{beta}$
        and the cost difference of such a proposed change as
            $\Delta = \texttt{new\_cost} - \texttt{cost}$,
    the \texttt{Accept} function calculates the probability of accepting a proposed change as
    \begin{equation}
        P_\textrm{accept} =
            \begin{cases}
                e^{-\beta \Delta}   &\Delta > 0     \\
                1                   &\Delta \le 0
                .
            \end{cases}
    \end{equation}
    The probability of accepting ``bad'' proposed changes where the cost increases (i.e., where $\Delta > 0$) approaches zero as $\beta$ increases.

\subsection{Compilation of time-evolution unitaries}\label{sec:appendix-stoq-compilation-time-evolution}

\begin{figure*}
    \centering
    \includegraphics[width=\linewidth]{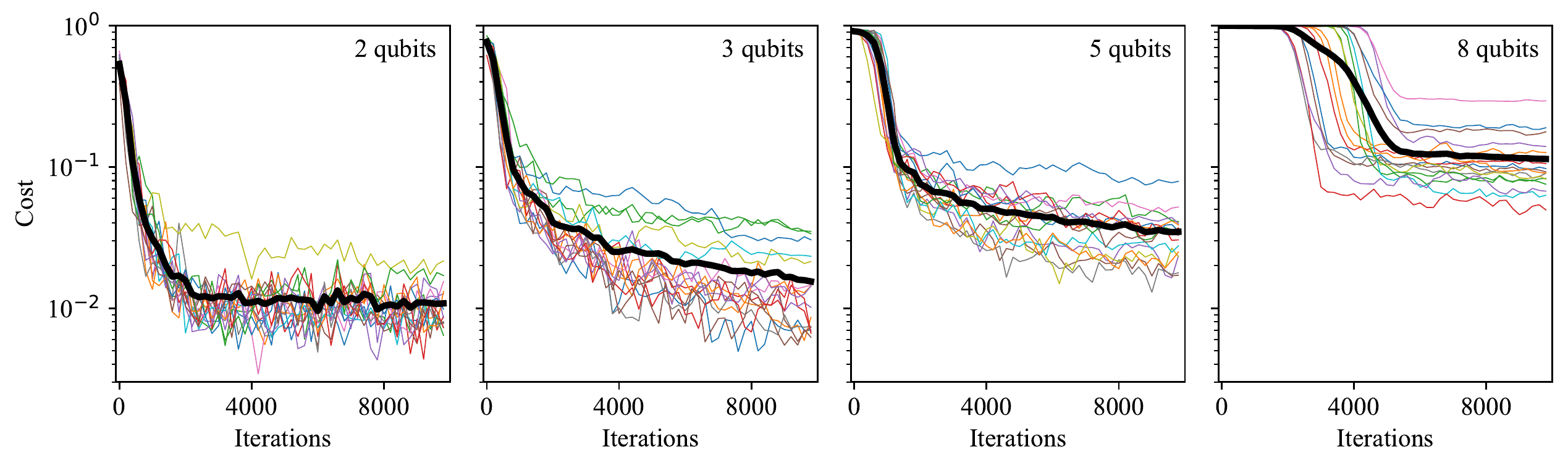}
    \caption{
        Compilation via STOQ for two-qubit, three-qubit, five-qubit, and eight-qubit versions of the time-evolution unitary
            from \equationref{eq:time-evolution-unitary}.
        Each of the 16 thin curves shows the value of the cost function from \equationref{eq:cost}
            during a single compilation using 10,000 iterations.
        The thick curve is the average of all runs.
    }
    \label{fig:time-evolution-unitary-compilation}
\end{figure*}

To demonstrate one possible (although not necessarily useful) application of STOQ,
    we choose an Ising-type Hamiltonian
        with nearest-neighbor coupling and transverse field
    \begin{equation}\label{eq:hamiltonian}
        H = \sum_{<i,j>} J_{ij} \sigma_x^{(i)}\sigma_x^{(j)}
            + \sum_{i} h_i \sigma_y^{(i)}
    \end{equation}
    where the coefficients $J_{ij}$ and $h_i$ are energies with particular values for each system size, as shown in \tableref{tab:coefficient-values}.

\begin{table}
    \centering
    \setlength\tabcolsep{5pt}
    \begin{tabular}{c | c | c | c | c | c | c | c | c | c | c | c | c | c | c | c}
        $n$ & $J_{12}$ & $J_{23}$ & $J_{34}$ & $J_{45}$ & $J_{56}$ & $J_{67}$ & $J_{78}$ & $h_1$ & $h_2$ & $h_3$ & $h_4$ & $h_5$ & $h_6$ & $h_7$ & $h_8$ \\
        \hline
        2   & 1.27     &          &          &          &          &          &           & 1.54  & 1.19  &       &       &       &       &       &       \\
        3   & 1.81     & 1.27     &          &          &          &          &           & 1.54  & 1.19  & 0.53  &       &       &       &       &       \\
        5   & 1.20     & 1.40     & 1.60     & 1.80     &          &          &           & 1.60  & 1.30  & 1.00  & 0.70  & 0.40  &       &       &       \\
        8   & 1.20     & 1.30     & 1.40     & 1.50     & 1.60     & 1.70     & 1.80      & 1.40  & 1.10  & 0.80  & 1.00  & 1.20  & 1.50  & 1.70  & 1.30  \\
    \end{tabular}
    \caption{\label{tab:coefficient-values}
        Coefficients used for application of STOQ to the
            $n$-qubit Ising model Hamiltonian in \equationref{eq:hamiltonian}.
            Values are energies in kHz where $\hbar = 1$.
    }
\end{table}

We then define the time-evolution unitary as $U_t(\tau) = e^{i H \tau}$, where we choose units such that $\hbar = 1$, and we concretely choose $\tau = 0.5 \textrm{ ms}$, such that
    \begin{equation}\label{eq:time-evolution-unitary}
        U = U_t(0.5 \textrm{ ms}) = e^{i H (0.5 \textrm{ ms})}
    \end{equation}
    is the target unitary for compilation.

To apply STOQ, we need also to choose a parameterized instruction set $G$ from which to approximately compile a sequence.
In a physical device, it is often the case that the dynamics are implemented such that each term in $H$ can be individually controlled.
To define $G$ for such a device, we express the Hamiltonian as 
        $H = \sum_k H_k$,
    where each $H_k$ is one of the
        $\sigma_x\sigma_x$ or $\sigma_y$
        terms from \equationref{eq:hamiltonian},
    and choose
    \begin{equation}
        G = \bigcup_k \left\lbrace e^{i H_k t} \right\rbrace
        \quad -\epsilon \tau \le t \le \epsilon \tau
    \end{equation}
    where the allowed range for $t$ is chosen such that the duration of each instruction is relatively short in comparison to the timescale of the dynamics of $H$. (In this demonstration we use $\epsilon = 0.2$.) 
    Negative times correspond to reversing the sign of the coefficient of a given term.
    (We note that for a general Hamiltonian, it is unlikely that each $H_k$ term is part of the native gate set of the device. In this case, $G$ should instead represent the interactions which are implemented natively on the device.)
    
We then apply STOQ to compile many sequences that approximately implement $U$, using two-qubit, three-qubit, five-qubit, and eight-qubit versions of the corresponding Hamiltonian. 
    \figureref{fig:time-evolution-unitary-compilation} reports the cost for 16 such compilations as a function of the number of iterations.
    (Each run of 10,000 iterations for the five-qubit system takes around 15 minutes to complete on a typical desktop computer.)
    We observe that the stochastic search process rapidly reduces the cost at first before noticeably leveling off. For the two-qubit and three-qubit systems, this cost approaches a limit near $10^{-2}$ after 10,000 iterations. For the larger systems, the final average cost is higher, although even for the eight-qubit system, the final cost reaches a value below $10^{-1}$ for some compilations.

\begin{figure*}
    \centering
    \includegraphics[width=\linewidth]{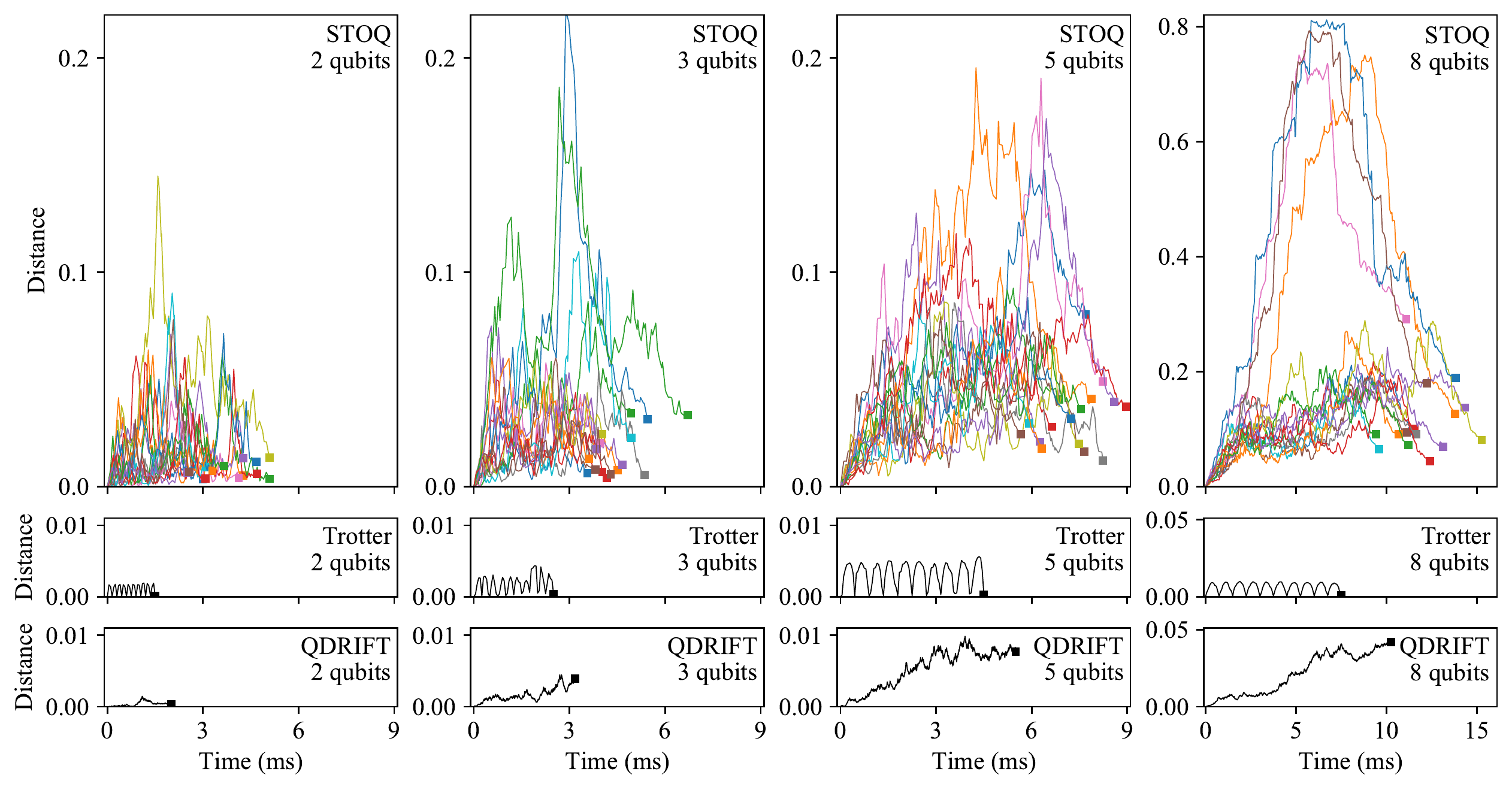}
    \caption{
        Distance from ideal path to compiled path,
            as defined in \equationref{eq:path-distance},
            for the time-evolution unitary from \equationref{eq:time-evolution-unitary},
            illustrating ``how different'' a particular compilation is from the path that would be followed by the ideal time evolution of the full Hamiltonian.
        The distance from the horizontal axis quantifies the deviation in the system state from the ideal simulation when executing the specified compilation.
        The ideal simulation would be a flat curve along the horizontal axis from $\tau=0.0 \textrm{ ms}$ to $\tau=0.5 \textrm{ ms}$.
            Results are shown for 2-qubit, 3-qubit, 5-qubit, and 8-qubit
                implementations of the Ising model Hamiltonian from
                \equationref{eq:hamiltonian}.
            Each curve represents the execution of one compiled sequence.
            Filled squares are used to plot the
                overall running time of the compiled sequence and
                final cost of each compilation.
        Top row depicts the execution of 16 independent STOQ compilations,
            each using 10,000 iterations.
            Each curve corresponds to a curve of the same color
                in \figureref{fig:time-evolution-unitary-compilation}.
        Middle row depicts the execution of a typical randomized Suzuki-Trotter compilation using 10 steps.
        Bottom row depicts the execution of a typical QDRIFT compilation using 1,000 repetitions. The horizontal axis of each plot represents the execution time of the compiled circuits.
    }
    \label{fig:compilation-distance}
\end{figure*}

To compare STOQ to existing compilation techniques, we also compile sequences to approximately implement $U$ using
        the randomized Suzuki-Trotter decomposition \cite{Childs2019FasterRandomization}
        and the QDRIFT stochastic compilation protocol \cite{Campbell2019RandomSimulation}.
    STOQ is designed to create more randomness in the resulting path taken through state space.
        To compare these paths quantitatively, we choose to compare the various methods to an ideal version where $H$ is directly implemented for time $\tau$.
        We define the \textit{ideal path} as the path taken by this ideal time evolution, and we define the \textit{compiled path} as the path taken by the compiled sequence, which we represent as a sequence of instructions $\{ G_1, \dots, G_M \}$.
        We then calculate the path distance $d_m$ from the ideal path to step $m$ of the compiled path, where $1 \le m \le M$, as
        \begin{equation}\label{eq:path-distance}
            d_m = \min_{t\,\in\,[0, \tau]}
                D_{\textrm{HS}} \left(
                    e^{i H t}, \ G_m G_{m-1} \cdots G_1 \right)
            ,
        \end{equation}
        where $D_{\textrm{HS}}$ is the distance metric defined in \equationref{eq:hilbert-schmidt-distance}. Thus $d_m$ is the shortest distance from step $m$ of the compiled path to any point in the ideal path.

Results for each compilation technique are shown
    in \figureref{fig:compilation-distance},
    and statistics for the five-qubit example
    are displayed in \tableref{tab:compilation-distance}.
We observe that the STOQ compilations result in a significantly greater path distance from the ideal evolution than the other approaches,
    and that the total running time of the compiled sequence resulting from the various compilations is within a factor of two.

However, the final cost of the STOQ compilations is typically at least an order of magnitude larger than the compilations created using the randomized Suzuki-Trotter and QDRIFT techniques, both of which can reach arbitrarily low costs by increasing the number of steps. This implies that STOQ would not be a useful tool for applications that require high-fidelity compilations.

\begin{table}
    \centering
    \begin{tabular}{l | c | c | c | c}
                  & Ideal        & Trotter   & QDRIFT   & STOQ    \\ \hline
        (a) Execution time (ms)      & 0.50         & 4.50      & 5.50     & 7.32    \\
        (b) Average distance & \textemdash  & 0.0032    & 0.0053   & 0.0469  \\
        (c) Maximum distance  & \textemdash  & 0.0056    & 0.0099   & 0.1133  \\
        (d) Final cost      & \textemdash  & 0.0003    & 0.0077   & 0.0328
    \end{tabular}
    \caption{\label{tab:compilation-distance}
        Statistics resulting from various compilations of the five-qubit time-evolution unitary
            from \equationref{eq:time-evolution-unitary},
        where the ideal evolution occurs for $\tau=0.5 \textrm{ ms}$.
        For each of the compilation techniques, means are listed for each of the following quantities:
            (a) total execution time of the compiled sequence,
            (b) average distance $\sum_{m=1}^M d_m$,
            (c) maximum distance $\max_{m\in[1,M]} d_m$,
            and (d) final cost $d_M$.
        Corresponds to five-qubit plots in \figureref{fig:compilation-distance}.
    }
\end{table}

\subsection{Compilation of random unitaries}\label{sec:appendix-stoq-compilation-random-unitaries}

\begin{figure*}
    \centering
    \includegraphics[width=0.99\linewidth]{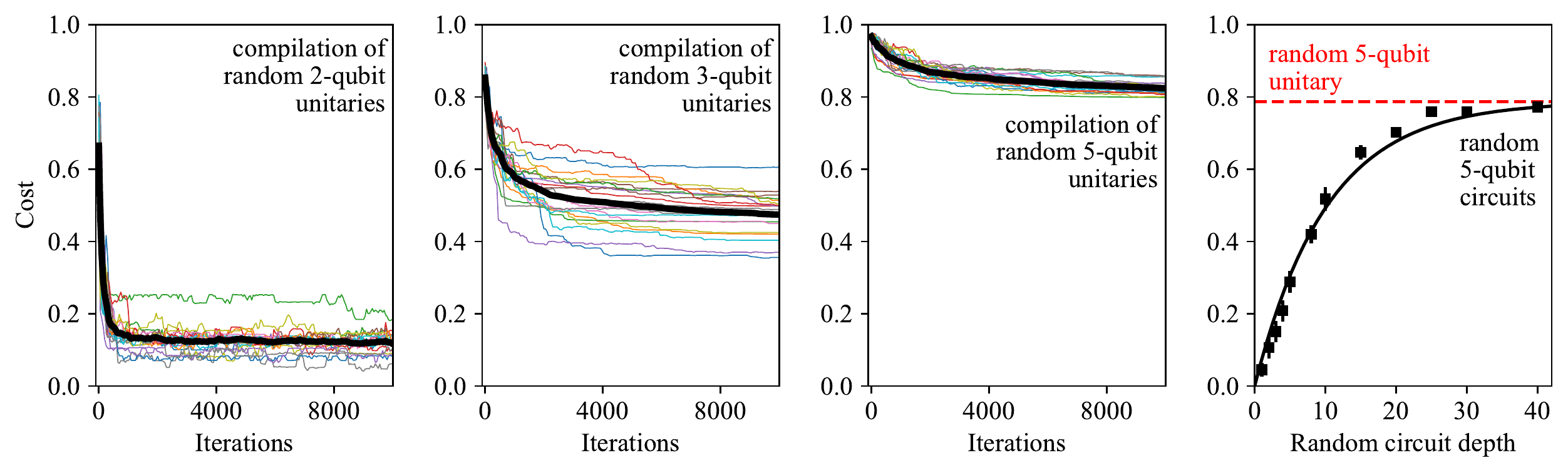}
    \caption{
        Compilation via STOQ of unitaries chosen uniformly at random from the Haar measure.
        The left three plots show the cost during the STOQ compilation process for randomly-generated 2-qubit, 3-qubit, and 5-qubit target unitaries.
        Each of the 20 thin curves shows the value of the cost function from \equationref{eq:cost}
            during a single compilation using 10,000 iterations.
            The thick curve is the average of all runs.
        The rightmost plot shows the final cost of the STOQ compilation for target unitaries generated by creating random 5-qubit circuits of varying average circuit depth.
            Circuit depth is calculated as the total number of instructions divided by the number of qubits.
            Each point is the average of 20 compilations using 100,000 iterations.
            Error bars indicate standard error of the mean.
            The solid line is an exponential decay fit with one free parameter.
            The dashed line represents the average final cost of compiling a randomly-generated 5-qubit unitary.
            Note that one would expect a similar scaling with circuit depth for general quantum circuits, regardless of whether they are generated randomly.
    }
    \label{fig:random-unitary-compilation}
\end{figure*}

In addition to being used for sparse or highly structured unitaries such as those generated from Hamiltonian time-evolution,
   the STOQ protocol can also be used to compile sequences that approximately implement purely random unitaries in terms of an arbitrary instruction set, without having any prior knowledge of the structure of the unitary.
Such an application of STOQ is not necessarily useful in general, but is included here for illustrative purposes.
   
\figureref{fig:random-unitary-compilation} shows
   typical results of repeatedly using the STOQ protocol
   to compile sequences for random two-qubit, three-qubit, and five-qubit unitaries,
   chosen uniformly at random from the Haar measure \cite{Mezzadri2006HowGroups},
   using a simple universal instruction set
     $G = \{ R(\theta, \varphi), XX(\theta) \}$.
   $R(\theta, \varphi)$ is a parameterized single-qubit rotation
      \begin{equation}
          \setlength\arraycolsep{2pt}
          R(\theta, \varphi) = 
            \begin{bmatrix}
           \cos{\frac{\theta}{2}} & -i e^{i\varphi} \sin{\frac{\theta}{2}} \\[2mm]
           i e^{-i\varphi} \sin{\frac{\theta}{2}} & \cos{\frac{\theta}{2}}
            \end{bmatrix}
      \end{equation}
      with $0 \le \theta < 2\pi$ and $0 \le \varphi < 2\pi$.
      $XX(\theta)$ is a parameterized two-qubit entangling gate
      \begin{equation}
          \setlength\arraycolsep{2pt}
          XX(\theta) = 
            \begin{bmatrix}
                \cos \theta         & 0         & 0         & -i \sin \theta  \\
                0         & \cos \theta         & -i \sin \theta    & 0 \\
                0         & -i \sin \theta         & \cos \theta         & 0  \\
                -i \sin \theta         & 0         & 0         & \cos \theta 
            \end{bmatrix}
      \end{equation}
      with $0 \le \theta < 2\pi$.
      We note that the instruction set $G$ is a typical native gate set that can be implemented by trapped-ion quantum devices.

We observe that the final costs of compilation of these random unitaries are significantly larger than for compilation of the time-evolution unitaries discussed in \appendixref{sec:appendix-stoq-compilation-time-evolution}.
In particular,
    the final cost is approximately 0.1 for two-qubit random unitaries, 0.5 for three-qubit random unitaries, and 0.8 for five-qubit random unitaries.
    This indicates that the quality of the approximation for such random unitary compilations scales poorly with system size.
    This is not surprising, since reaching the vast majority of states in the Hilbert space of a system requires circuits of depth which grows exponentially with the dimension of the Hilbert space \cite{Knill1995ApproximationCircuits, Poulin2011QuantumSpace}.
    Nonetheless, the compilations generated by this method may be useful in scenarios where high-fidelity approximations are not required.
    
We also observe that the final cost of such random unitary compilations is relatively stable over a wide range of STOQ parameter values.
    Two primary parameters that can be adjusted in the STOQ algorithm in \figureref{fig:stoq-algorithm} are
        the annealing rate $\Delta\beta$,
            which is used to increment $\beta$ at each step inside the \texttt{IncreaseBeta} function,
        and the probability $p_\textrm{append}$ that the search appends an instruction (as opposed to removing an instruction) at each step,
            which occurs inside the \texttt{RandomChange} function.
    For compilation of three-qubit random unitaries,
        and for values $\Delta\beta \in \{0.001, 0.01, 0.1, 0.5\}$ and $p_\textrm{append} \in \{0.2, 0.5, 0.8\}$,
        we find that the average final cost remains between 0.398 (for $\Delta\beta=0.5$ and $p_\textrm{append}=0.2$) and 0.448 (for $\Delta\beta=0.001$ and $p_\textrm{append}=0.5$),
        where each pair of parameter values is averaged over 32 compilations using 100,000 iterations each.

To provide insight into the low-fidelity approximations of random unitaries produced by STOQ, we consider the case of target unitaries generated by random circuits of varying depth.
    To do this, we generate random five-qubit circuits of average depth ranging from 1 to 40, where the average depth is calculated as the total number of instructions divided by the number of qubits.
    The rightmost plot in \figureref{fig:random-unitary-compilation} shows the final compilation cost after applying STOQ to generate an approximate compilation of the unitary corresponding to each random circuit.
    As might be expected, we observe that STOQ generates relatively high-fidelity approximations for shallow circuits, since such unitaries are known to be reachable with a fixed number of instructions. But as the circuit depth increases, the resulting unitaries begin to look more like random unitaries, and the final compilation cost approaches that of the randomly-generated five-qubit unitary discussed previously.
    Indeed, we expect a similar scaling with circuit depth for quantum circuits in general, regardless of whether they are randomly generated, since the size of the reachable state space grows exponentially with the depth of the circuit.

\subsection{Discussion}\label{sec:appendix-stoq-discussion}

We note that because the STOQ protocol requires calculating the product of the compiled sequence during each iteration, the computational cost of each iteration grows exponentially in the system size $n$. Therefore, STOQ is unlikely to be useful for system sizes of more than around 10 qubits.
In particular, for compilation of time-evolution unitaries, this clearly means that STOQ will be less computationally efficient when compared to compilation methods based on product formulas,
which in general have a computational cost   that depends only on the number of terms in the Hamiltonian and is independent of the system size.

We note that unitaries generated via time evolution of a Hamiltonian often benefit from the sparsity of the Hamiltonian.
    In general, an $n$-qubit Hamiltonian has $4^n$ coefficients when expressed in the basis of Pauli operators.
    For the five-qubit version of the Hamiltonian in \equationref{eq:hamiltonian}, only nine of these 1024 coefficients are non-zero.
    Sparsity in the Hamiltonian greatly limits the subspace of the full operator space that can be reached by via time evolution, which in turn makes compilation a more feasible task and
    allows techniques such as Suzuki-Trotter and QDRIFT to be highly efficient.
Because the number of possible step proposals during each iteration of the STOQ search process is determined by the number of terms in the Hamiltonian, it is reasonable to infer that STOQ is similarly more effective when the problem structure contains such sparsity.
    This is further evidenced by the inability of the STOQ protocol to efficiently obtain low cost values when compiling sequences for random target unitaries, which are not sparse in general.

As demonstrated, STOQ has some features that are distinct from other methods.
One notable feature is the capability of generating results with arbitrary gate sets, regardless of whether the gates are fixed or parameterized or whether the gate set is universal.
In addition, repeated application of STOQ provides many independent approximate compilations of the same unitary, and each compilation creates a sequence that will cause the system state to traverse a different path in state space. As depicted in \figureref{fig:compilation-distance}, even stochastic techniques such as randomized Suzuki-Trotter or QDRIFT result in a compiled sequence that will cause the system state to follow very nearly the same path in state space as the deterministic version, whereas sequences generated by STOQ cause the system to traverse unique paths that can differ greatly from the ideal path and from each other.

\end{document}